\documentclass[aps,twocolumn,superscriptaddress,longbibliography]{revtex4-1}

\usepackage{bm,hyperref}
\usepackage{amsmath}
\usepackage{hyperref}
\hypersetup{
    colorlinks=true,
    linkcolor=blue,
    citecolor=black,
    filecolor=black,
    urlcolor=blue}
\makeatletter
\renewcommand*\eqref[1]{\hyperref[{#1}]{\textup{\normalfont(\ref*{#1})}}}
\makeatother

\usepackage{physics}
\usepackage{siunitx}
\usepackage{array}
\usepackage{bbold}
\usepackage{tabularx}
\usepackage{cleveref}
\usepackage{natbib} 

\newcommand{\upperRomannumeral}[1]{\uppercase\expandafter{\romannumeral#1}}

\newcommand{\svdots}{\raisebox{3pt}{\scalebox{.6}{$\vdots$}}}
\newcommand{\sddots}{\raisebox{3pt}{\scalebox{.6}{$\ddots$}}}

\usepackage[]{graphicx}
\usepackage{caption}
\usepackage{subcaption}

\usepackage{tikz}
\usetikzlibrary{intersections,decorations.pathreplacing,decorations.markings,calc,angles,quotes,arrows.meta,patterns}
\usepackage[prependcaption,textsize=tiny,textwidth=3cm]{todonotes}
\graphicspath{ {images/} }

\usepackage{mathrsfs}
\usepackage{dsfont}

\begin{document}
\title{Quantum optics of chiral and antichiral waveguide arrays}

\author{Peng Wang}
\affiliation{Department of Applied Physics, Yale University, New Haven, CT 06511, USA}

\author{Erik Orvehed Hiltunen}
\affiliation{Department of Mathematics, University of Oslo, Norway}

\author{John C. Schotland}
\affiliation{Department of Mathematics, Yale University, New Haven, CT 06511, USA}
\affiliation{Department of Physics, Yale University, New Haven, CT 06511, USA}

\begin{abstract}
We study single-photon scattering by atoms in arrays of one-way waveguides. We investigate both chiral and antichiral arrays, where the one-way waveguides are aligned in the same and opposite directions, respectively. In the chiral array, reciprocity is broken: one of the (spatial) dimensions is time-like, resulting in a light-cone feature of the scattered fields. In contrast, the antichiral array preserves reciprocity and exhibit scattering behavior typical of wave systems. In analogy with classical physical optics, we exmaine the geometrical optics, diffraction, and scattering regimes in the waveguide arrays. We illustrate our results using numerical simulations.
\end{abstract}

\maketitle
\section{Introduction}
Waveguide quantum electrodynamics has attracted considerable recent interest, providing a versatile platform for investigating novel light-matter interactions \cite{roy2017colloquium, sheremet2023waveguide,gonzalez2024light,turschmann2019coherent}. Notable effects include modifications of spontaneous emission \cite{shen2005coherent, siampour2023observation,mitsch2014quantum}, strong photon-photon correlations \cite{PhysRevLett.98.153003,PhysRevLett.121.143601} and long-range interactions between atoms mediated by guided modes \cite{PhysRevResearch.2.043213,solano2017super,kim2018super}. The strong spatial confinement inherent to waveguides enhances coupling between propagating modes and quantum emitters, enabling precise control of radiative and coherent properties of the system. On the application side, waveguide systems offer substantial potential in quantum technologies ranging from single-photon detection to integrated photonic processors \cite{akhlaghi2015waveguide, politi2008silica, wang2020integrated}. In this regard, coupled waveguide arrays are of particular interest as a platform for realizing scalable quantum circuits and photonic quantum networks.

One-way waveguides, in which light propagation becomes unidirectional, have been proposed as promising carriers of quantum information \cite{lodahl2017chiral, PRXQuantum.6.020101, PhysRevLett.117.240501,PhysRevA.94.063817,PhysRevResearch.2.043048}. The violation of reciprocity in such waveguides can arise from an asymmetric dispersion relation or enhanced spin-orbit coupling of light at subwavelength scales \cite{petersen2014chiral,sollner2015deterministic,coles2016chirality}. A model of single photons in a one-way waveguide containing many two-level atoms was introduced in \cite{mirza2017chirality,mirza2018influence}. In more recent work, the transport properties of a one-dimensional array of one-way waveguides were investigated~\cite{hoskins2023quantum}. In a suitable continuum limit, the one-photon amplitude of a single-excitation state was shown to obey a partial differential equation. Depending on the alignment of the waveguides, different  models are obtained. Remarkably, if the waveguides are aligned in the same direction, which we refer to as a \emph{chiral} array, the effective equation takes the form of a (1+1)-dimensional Dirac equation, where the coordinate along the waveguides is time-like while the transverse coordinate is space-like. As a consequence, scattering in the chiral array is analogous to that of time-dependent materials, a topic attracts recent attention due to phenomena such as frequency conversion \cite{yin2022floquet}, parametric amplification \cite{cullen1958travelling, raiford1974degenerate} and Floquet topological insulators \cite{fleury2016floquet,rechtsman2013photonic}. In contrast, if the waveguides are aligned in alternating directions, which we refer to as an \emph{antichiral} array, both coordinates are space-like. The resulting scattering theory is similar to that in classical optics.

In this paper, we substantially expand upon the results reported in~\cite{hoskins2023quantum}. In particular, we compare and contrast the propagation of single photons in chiral and antichiral arrays across various physical settings, analogous to those arising in classical physical optics. We begin by considering geometrical optics, a regime that holds when the atomic density varies on scales much larger than the optical wavelength. From the Dirac equations, we derive the according eikonal and transport equations governing the geometrical optics in each array and solve them along characteristic curves that are known as rays in classical optics. Next we consider diffractive corrections to geometrical optics and derive the corresponding diffraction integrals. To this end, we construct propagators for the Dirac equations by performing a Fourier transform in the transverse direction and illustrate the results in the case of diffraction by a single-slit aperture. Finally, we investigate the scattering of single photons from arbitrary atomic densities. This entails developing the scattering theory and related integral equations for chiral and antichiral Dirac equations. These results are then applied to discrete and continuous atomic densities. We note that scattering theory for Dirac equations has typically been studied in the context of high-energy electrons in Coulomb potentials \cite{parzen1950scattering, scheck1968approximate}, and has recently received significant attention as an effective model for hexagonally symmetric scattering problems in the vicinity of Dirac points \cite{fefferman2014wave,ammari2020high,wallace1947band, ablowitz2009conical,peres2009scattering}.
 


This paper is organized as follows. In \Cref{sec:setting}, we recall the quantum electrodynamics of chiral and antichiral waveguide arrays and derive the corresponding Dirac equations. We consider the high-frequency behavior of the Dirac equations in \Cref{sec:goptics}, which corresponds to the regime of geometrical optics.  In \Cref{sec:diff}, we study diffraction in waveguide arrays and derive propagators for the chiral and antichiral Dirac equations. In \Cref{sec:scattering}, we study the scattering theory for the Dirac equations. Using an integral equation approach, we investigate scattering from collections of single atoms. Scattering from a spherical obstacle is then considered in the antichiral array, for which explicit formulas for the scattered field are obtained. Finally, we study the transmission problem for a slab. The paper concludes with a discussion in \Cref{sec: discussion}. Details of the calculations are presented in the appendices.



\section{Chiral and antichiral arrays}
\label{sec:setting}
\begin{figure}[t]
\includegraphics[width=\linewidth]{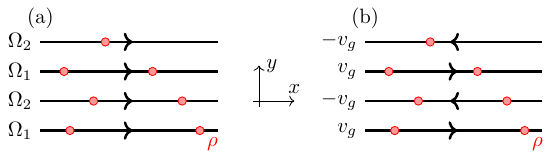}

\caption{Ilustrating the chiral (a) and antichiral (b) arrays. The chiral array is composed of one-way waveguides interspersed at alternating frequencies $\Omega_1$ and $\Omega_2$. The antichiral array is composed of one-way waveguides with alternating group velocities. In both cases, atoms are distributed within the array with number density $\rho$. }\label{fig:arrays}
\end{figure}

In this section, we consider the quantum electrodynamics of a one-dimensional lattice of one-way waveguides. Each waveguide is assumed to contain many two-level atoms. Throughout this work, we will be concerned with  two distinct geometries as shown in \Cref{fig:arrays}:
the {chiral} array, which comprises two interpenetrating sublattices with alternating waveguide frequencies, and the {antichiral} array, which consists of two sublattices with opposite group velocities. 

\subsection{Model}
We begin by recalling the model introduced in~\cite{hoskins2023quantum}. The total Hamiltonian of the system is of the form
\begin{equation}
    H=H_A+H_F+H_I,
\end{equation}
where $H_A$, $H_F$ and $H_I$ denote the atomic, optical field and interaction Hamiltonians, respectively. 
We adopt a real-space quantization procedure, in which the atoms and the optical field are treated on the same footing. To this end, the atomic Hamiltonian $H_A$ is given by  
\begin{equation}
    H_A=\hbar\omega_0\int dx\sum_n\rho_n(x)\sigma_n^{\dagger}(x)\sigma_n(x) .
\end{equation}
Here $\omega_0$ is the atomic transition frequency and $\rho_n(x)=\sum_{x_n}\delta(x-x_n)$ is the number density of atoms in the $n$th waveguide, where the sum is the over the positions $y_n$ of the atoms in the waveguide. In addition, $\sigma_n^{(\dag)}(x)$ is the lowering (raising) operator for an atom at position $x$ in waveguide $n$, which obeys the anticommutation relations
\begin{equation}
    \{\sigma_n(x),\sigma^{\dagger}_n(x)\}= 1 \ , \quad \{\sigma_n(x),\sigma_n(x)\} = 0
    \label{eq:anti_commute}
\end{equation}
and the commutation relations
\begin{equation}
    [\sigma_n(x),\sigma^{\dagger}_m(y)]= 0 \ , \quad [\sigma_n(x),\sigma_m(y)] = 0 \ ,
    \quad x\neq y \ \text{and} \ n\neq m \ .
    \label{eq:commute}
\end{equation}
That is, the atomic operators anticommute for an atom and commute for different atoms. The mixed commutation relations prohibit the double excitation of an atom while allowing the transfer of an excitation from one atom to another.

The field Hamiltonian $H_F$ is given by
\begin{multline}
    H_F=\hbar\int dx\sum_n\Big(\phi^{\dagger}_n(x)(\Omega_n-\mathrm{i}v_n\partial_x)\phi_n(x)\\+J_0[\phi^{\dagger}_n(x)\phi_{n+1}(x)+\phi^{\dagger}_{n+1}(x)\phi_n(x)]\Big),
\end{multline}
where $v_n$ is the group velocity of the $n${th} waveguide. Here, we suppose that the dispersion relation of the waveguide is linearized around a frequency  $\Omega_n$ and nearest-neighbor waveguides are coupled through evanescent waves with constant coupling strength $J_0$. The field operator $\phi_n^{(\dag)}(x)$, which annihilates (creates) a photon at position $x$ in waveguide $n$, obeys the commutation relations
\begin{equation}
    [\phi_n(x),\phi^{\dagger}_m(y)] =\delta_{nm}\delta(x-y) \ , \quad [\phi_n(x),\phi_m(y)] = 0 .
    \label{eq:commute}
\end{equation}
The interaction between the atoms and the field is described by the Hamiltonian
\begin{equation}
    H_I=\hbar g\int dx\sum_n\rho_n(x)[\sigma^{\dagger}_n(x)\phi_n(x)+\sigma_n(x)\phi^{\dagger}_n(x)],
\end{equation}
where the constant $g$ measures the interaction strength, and we have made the rotating wave approximation.

The stationary behavior of the system is governed by the time-independent Schrödinger equation
\begin{equation}
H|\Psi\rangle=E_0|\Psi\rangle
\label{eq:schrodinger} ,
\end{equation}
where $|\Psi\rangle$ is the state of the system and $E_0$ is the energy.
We suppose that the system is in a single-excitation state of the form
\begin{equation}
    |\Psi\rangle=\int dx\sum_n\left(a_n(x)\rho_n(x)\sigma^{\dagger}_n(x)+\psi_n(x)\phi^{\dagger}_n(x)\right)|0\rangle,
\label{eq:state_expr}
\end{equation}
where $a_n(x)$ and $\psi_n(x)$ are the corresponding probability amplitudes of exciting an atom and finding a photon at position $x$ in waveguide $n$. By inserting \eqref{eq:state_expr} into \eqref{eq:schrodinger} and utilizing the anticommutation relation \eqref{eq:anti_commute} and the commutation relation \eqref{eq:commute}, we obtain the equations obeyed by $a_n(x)$ and $\psi_n(x)$:
\begin{equation}
    -\mathrm{i}\hbar v_n\partial_x \psi_n+\hbar\Omega_n\psi_n+\hbar J_0(\psi_{n-1}+\psi_{n+1})+ V_n(x)\psi_n=E_0\psi_n,
    \label{eq:field_amplitude_discrete}
\end{equation}
\begin{equation}
    a_n(x)=\frac{\hbar g}{E_0-\hbar \omega_0}\psi_n(x),
\end{equation}
where the potential $V_n(x)$ is defined by
\begin{equation}
    V_n(x)=\frac{\hbar^2 g^2}{E_0-\hbar \omega_0}\rho_n(x).
\end{equation}

\subsection{Dirac equations}
We now derive the Dirac equations for chiral and antichiral waveguide arrays. Recall that each array is composed of two interpenetrating subarrays with different types of waveguides. We use $v_{1,2}$ and $\Omega_{1,2}$ to denote the corresponding group velocities and waveguide frequencies, respectively. In addition, we define the frequency difference $\Omega=(\Omega_1-\Omega_2)/2$ and the energy $E=E_0-\hbar(\Omega_1+\Omega_2)/2$. With these definitions, we split \eqref{eq:field_amplitude_discrete} into even and odd parts by modifying the amplitudes according to
 \begin{equation}
    \begin{split}
        &\Tilde{\psi}_1(x,n)=(-1)^n\psi_{2n}(x),\\
        &\Tilde{\psi}_2(x,n)=-\mathrm{i}(-1)^n\psi_{2n+1} .
    \end{split}
\end{equation}
Eq.~\eqref{eq:field_amplitude_discrete} thus becomes
\begin{equation}
\begin{split}
-\mathrm{i}\hbar v_1\partial_x\Tilde{\psi}_1+\hbar\Omega\Tilde{\psi}_1+\mathrm{i}\hbar J_0\partial_n\Tilde{\psi}_2 +V_{2n}\Tilde{\psi}_1 &=E\Tilde{\psi}_1 ,\\    
-\mathrm{i}\hbar v_2\partial_x\Tilde{\psi}_2 -\hbar\Omega\Tilde{\psi}_2 +\mathrm{i}\hbar J_0\partial_{n+1}\Tilde{\psi}_1(x,n+1)+V_{2n+1}\Tilde{\psi}_2 &=E\Tilde{\psi}_2 ,
\end{split}\label{eq:discrete_dirac}
\end{equation}
where we have introduced the difference operator $\partial_{n}f(n) = f(n)-f(n-1)$. 
In the continuum limit, where the spacing between the waveguides $\ell\to 0$, \eqref{eq:discrete_dirac} becomes
\begin{equation}
    \begin{split}
-\mathrm{i}\hbar v_1\partial_x\psi_1+\hbar\Omega\psi_1 +\mathrm{i}\hbar J\partial_y\psi_2 +U \psi_1 &=E\psi_1 , \\    
-\mathrm{i}\hbar v_2\partial_x\psi_2 -\hbar\Omega\psi_2 +\mathrm{i}\hbar J\partial_y\psi_1 +U \psi_1 &=E\psi_2  ,
\end{split}\label{eq:continuous_dirac}
\end{equation}
where we have introduced the transverse coordinate $y$ to replace the waveguide index $n$ and defined $J=J_0\ell$. We have also
redefined the field amplitude and potential according to $\Tilde{\psi}_{1,2}(x,n)\rightarrow\psi_{1,2}(x,y)$ and $V_n(x) \rightarrow U(x,y)$, respectively.

Eq.~\eqref{eq:continuous_dirac} can be recast as a Dirac equation. To proceed, we introduce a two-dimensional  vector field $\psi=(\psi_1,\psi_2)^T$.
The case of a chiral array corresponds to taking equal group velocities $v_1=v_2=v_g$ and $\Omega \neq 0$ in each subarray. We find that $\psi$ obeys the Dirac equation
\begin{equation}
-\mathrm{i}\hbar v_g\partial_{x}\psi+\mathrm{i}\hbar J\alpha\partial_{y}\psi+\hbar\Omega\beta\psi+U(x,y)\psi=E\psi,\label{eq:dirac_chiral}
\end{equation}
where $\alpha$ and $\beta$ are the Pauli matrices 
\begin{equation}
    \alpha=\begin{pmatrix}
        0 &1\\
        1 &0
    \end{pmatrix}, \quad 
    \beta=\begin{pmatrix}
        1&0\\
        0&-1
    \end{pmatrix}.
\end{equation}
We note that the oscillatory term in \eqref{eq:dirac_chiral} can be removed by making the transformation $\psi\rightarrow\psi e^{\mathrm{i}Ex/(\hbar v_g)}$, so that  \eqref{eq:dirac_chiral}  becomes
\begin{equation}
    -\mathrm{i}\hbar v_g\partial_{x}\psi+\mathrm{i}\hbar J\alpha\partial_{y}\psi+\hbar\Omega\beta\psi+U(x,y)\psi =0.
\label{eq:dirac_chiral_1}
\end{equation}
It will prove useful to rescale the $y$ coordinate according to $y\to J/v_g y$ and define the wavenumber $k_c=\Omega/v_g$ and the potential $V_c=U/(\hbar\Omega)$. 
Eq.~\eqref{eq:dirac_chiral_1} then takes the form
\begin{equation}
   -\mathrm{i}\partial_{x}\psi+\mathrm{i}\alpha \partial_{y}\psi+k_{c} (\beta +V_c(x,y))\psi=0 , \label{eq:chiral_dirac_nondim}
\end{equation}
where the identity matrix that would normally appear in front of $V_c$ is omitted for notational simplicity.

The case of an antichiral array assumes that the group velocities alternate in sign with $v_1=-v_2=v_g$ and $\Omega= 0$ in each subarray. Consequently, the vector field $\psi$ given by \eqref{eq:continuous_dirac} obeys the Dirac equation
\begin{equation}
\mathrm{i}\beta\partial_{x}\psi+\mathrm{i}\alpha\partial_y\psi+k_a (V_a(x,y)-1)\psi=0,
\label{eq:anti_dirac_nondim}
\end{equation}
where $k_a=\Omega/(\hbar v_g)$, $V_a=U/E$, and the $y$ coordinate has been rescaled as above.

 
 We now make some observations regarding the Dirac equations. 
We note that Eq.~\eqref{eq:chiral_dirac_nondim} is a (1+1)-dimensional Dirac equation, where the coordinate $x$ is time-like, the coordinate $y$ is space-like and $\Omega$ plays the role of the mass. Meanwhile, it is a hyperbolic partial differential equation (PDE), admitting wave-like solutions with propagating singularities. Thus, the smoothness of the solutions depends on that of the initial and boundary conditions. In contrast, Eq.~\eqref{eq:anti_dirac_nondim} is a (2+0)-dimensional Dirac equation, with both coordinates being space-like. Accordingly, it is an elliptic PDE. Equations of this type have smooth solutions, which are independent of the smoothness of the boundary conditions. The distinction between the time-like and space-like character of the Dirac equations has important implications, which were explored in numerical simulations in \cite{hoskins2023quantum}. In this paper, we present a detailed analysis that explains and expands upon the numerical results. The analysis is carried out in three regimes, associated with the phenomena of geometrical optics, diffraction, and scattering. 

\subsection{Conservation of probability}
\label{sec:cons}
We next derive the conservation laws governing the field $\psi$. The starting point is the Dirac equations \eqref{eq:chiral_dirac_nondim} and \eqref{eq:anti_dirac_nondim}. We begin with the chiral case \eqref{eq:chiral_dirac_nondim}. Multiplying \eqref{eq:chiral_dirac_nondim} by $\psi^{\dagger}$  from the left and subtracting the result from its hermitian conjugate, we obtain
\begin{equation}
    \partial_{x}(\psi^{\dagger}\psi)-\partial_{y}(\psi^{\dagger}\alpha\psi)=0.\label{eq:chiral_conserve_1}
\end{equation}
If we define the probability current as
\begin{equation}
    \bm{j}= \psi^{\dagger}\psi\hat{\bm{x}}-\psi^{\dagger}\alpha\psi \hat{\bm{y}},
    \label{eq:chiral_current}
\end{equation}
\eqref{eq:chiral_conserve_1} can be rewritten in the form
\begin{equation}
    \nabla\cdot{\bm{j}}=0, \label{eq:current_conserv}
\end{equation}
which describes the conservation of probability.

In the antichiral case, we follow the same procedure as above and find that the current $\bm{j}$ shall be defined as
\begin{equation}
\bm{j}=\psi^{\dagger}\beta\psi\hat{\bm{x}}+\psi^{\dagger}\alpha\psi\hat{\bm{y}}
\label{eq:anti_current}
\end{equation}
to satisfy the conservation law 
of the form \eqref{eq:current_conserv}. 

\section{Geometrical optics}
\label{sec:goptics}
In this section, we develop the theory of waveguide arrays in the regime where the potentials $V_c$ and $V_a$ vary slowly compared to the characteristic wavelengths $2\pi/k_c$ and $2\pi/k_a$, respectively. This regime is analogous to geometrical optics, where the Dirac equations \eqref{eq:chiral_dirac_nondim} and \eqref{eq:anti_dirac_nondim} play the role of the wave equation for the optical field.



In this regime, we expect the solutions to \eqref{eq:chiral_dirac_nondim} and \eqref{eq:anti_dirac_nondim} should behave approximately as plane waves on scales that are large compared to the wavelength. To this end, we express the field $\psi$ in the form
\begin{equation}
    \psi(x,y)=e^{\mathrm{i}kS(x,y)}\phi(x,y) .
\label{eq:field_separate}
\end{equation}
Here $S$ is the eikonal, $\phi$ denotes the field amplitude and $k$ is the wavenumber given by $k_c$ in the chiral case and $k_a$ in the antichiral case, respectively. In both cases, we expand $\phi$ in powers of $1/k$ according to
\begin{equation}
\begin{split}
    \phi(x,y)&=\phi_{0}(x,y)+\frac{1}{k}\phi_{1}(x,y)+\cdots\\
    &=
    \begin{pmatrix}
    u_{0}(x,y)\\
    v_{0}(x,y)
    \end{pmatrix}+
    \frac{1}{k}
    \begin{pmatrix}
    u_{1}(x,y)\\
    v_{1}(x,y)
    \end{pmatrix}+\cdots.
\end{split}\label{eq:amplitude_expansion}
\end{equation}
We now substitute \eqref{eq:field_separate} and \eqref{eq:amplitude_expansion} into the Dirac equations \eqref{eq:chiral_dirac_nondim} and \eqref{eq:anti_dirac_nondim}. For the chiral array, we find that
\begin{equation}
    \mathcal A(S) \phi_0+\frac{1}{k_c}\bigl(\mathcal A(S) \phi_1+\mathrm{i}(-\partial_{x}\phi_0+\alpha\partial_{y}\phi_0)\bigr)+O\left(\frac{1}{k_c^2}\right)=0,
\label{eq:chiral_dirac_expansion}
\end{equation}
where the matrix $\mathcal A(S)$ is defined as
\begin{equation}
    \mathcal A(S)=(\partial_{x}S + V_c)\mathbb{1} -\alpha\partial_{y}S+\beta ,
\end{equation}
with $\mathbb 1$ being the identity matrix.
Upon collecting terms of the same order, we obtain
\begin{equation}
    \mathcal A(S)\phi_0=0 \label{eq:chiral_geomtric_order_1}
\end{equation}
at order $O(1)$ and
\begin{equation}
   \mathcal A\phi_{1}+\mathrm{i}(-\partial_{x}\phi_{0}+\alpha\partial_{y}\phi_{0})=0 
   \label{eq:chiral_geomtric_order_2}
\end{equation}
at order $O(1/k_c)$. We note that the determinant of the matrix $\mathcal A$ must be zero to allow for nontrivial solutions $\phi_0$. This condition yields the eikonal equation \begin{equation}
(\partial_{x}S+V_c)^{2}-(\partial_{y}S)^{2}=1.\label{eq:chiral_eikonal}
\end{equation}
It follows from \eqref{eq:chiral_geomtric_order_1} that the two components of $\phi_0$ obey
\begin{equation}
    (\partial_{x}S+V_c+1)u_0-(\partial_y S)v_0=0.\label{eq:chiral_transport_1}
\end{equation}
Moreover, by eliminating $\phi_1$ in \eqref{eq:chiral_geomtric_order_2} with the aid of \eqref{eq:chiral_eikonal}, we obtain a second equation for $\phi_0$:
\begin{equation}
    (\partial_{x}S+V_c -1)(-\partial_x u_0+\partial_{y} v_0)+\partial_{y}S(-\partial_x v_{0}+\partial_{y}u_{0})=0.\label{eq:chiral_transport_2}
\end{equation}
Eqs.~\eqref{eq:chiral_transport_1} and \eqref{eq:chiral_transport_2} constitute the transport equations for the chiral array. The eikonal and transport equations are the fundamental equations governing the geometrical optics of the chiral array.

We now follow the same approach for the antichiral array.  In this case, the eikonal and transport equations are, respectively,
\begin{equation}
    (\partial_{x}S)^2+(\partial_{y}S)^2=(V_a-1)^2,
    \label{eq:geo_eikonal_antichiral}
\end{equation}
and
\begin{align}
     &\bigl(\partial_xS-(V_a-1)\bigr)u_0+(\partial_yS)v_0=0,\\
    &\bigl(\partial_xS+(V_a-1)\bigr)(\partial_xu_0+\partial_yv_0)+\partial_yS(-\partial_xv_0+\partial_yu_0)=0.
\nonumber
\end{align}

The eikonal and transport equations can be solved by the method of characteristics, in which the PDEs are reduced to a system of ordinary differential equations (ODEs). We begin with the eikonal equation, which can be expressed in the form
\begin{equation}
    F(x,y,S,q,p)=0 ,
    \label{eq: general_eikonal_eq}
\end{equation}
for a suitably chosen function $F$.
Here $q$ and $p$ denote the partial derivatives $\partial_xS$ and $\partial_yS$, respectively. The characteristic curves $\bigl(x(t),y(t)\bigr)$, which we parameterize by $t$, are defined as the solutions of
\begin{equation}
     \frac{dx}{dt}=\partial_{q}F , \quad
     \frac{dy}{dt}=\partial_{p}F .
 \label{eq:general_eikonal_character}
\end{equation}
The characteristic curves are referred to as rays in geometrical optics.
It follows from \eqref{eq: general_eikonal_eq} and \eqref{eq:general_eikonal_character} that the eikonal $S$ and its partial derivatives $q$ and $p$  obey the ODEs
\begin{equation}
        \frac{dS}{dt}=p\partial_{p}F+q\partial_{q}F,\ \
        \frac{dq}{dt}=-\partial_{x}F-q\partial_{S}F,\ \
        \frac{dp}{dt}=-\partial_{y}F-p\partial_{S}F.
\label{eq:geo_general_ode}
\end{equation}
\textcolor{black}{Using \eqref{eq:general_eikonal_character} and \eqref{eq:geo_general_ode}, we find that the curvature of the characteristic curves can be expressed as
\begin{equation}
    \kappa_c = \frac{\partial_yV_c}{(1+2p^2)^{3/2}},\quad \kappa_a = \biggl|\frac{1}{V_1}\biggl(\nabla V_1 - \frac{1}{2V_1}\frac{dV_1}{dt}\hat{\bm{k}}\biggr)\biggr|
    \label{eq:curvature}
\end{equation}
in the chiral and antichiral arrays, respectively. Here, $\hat{\bm{k}}$ denotes the tangent direction of the curves and $V_1 = V_a-1$. We note that the curvature in the chiral array only depends on the $y$ component of the potential gradient, which is consistent with the fact that the $x$ coordinate is time-like with backscattering prohibited. In contrast, antichiral curvature bends toward the direction of the potential gradient in a similar manner of classical optics.
}
We proceed by noting that the transport equations along the characteristic curves can be reduced to the ODEs as
\begin{equation}
\begin{split}
    \frac{1}{u_0}\frac{du_0}{dt}&=-p\partial_xh_c+(q+V_c-1)\partial_yh_c,\\
    v_0&=h_c(x,y)u_0,
\end{split} \label{eq:chiral_geo_amplitude}
\end{equation}
in the chiral array and
\begin{equation}
\begin{split}
    \frac{1}{u_0}\frac{du_0}{dt}&=p\partial_xh_a-(q+V_a-1)\partial_yh_a,\\
    v_0&=h_a(x,y)u_0,
\end{split} \label{eq:antichiral_geo_amplitude}
\end{equation}
in the antichiral array, respectively, where the functions $h_c$ and $h_a$ are defined as
\begin{equation}
    h_c=\frac{V_c+q+1}{p},\quad h_a=\frac{V_a-q-1}{p}.
\end{equation}
We point out that the transport equations \eqref{eq:chiral_geo_amplitude} and \eqref{eq:antichiral_geo_amplitude} can be readily solved once the eikonal $S$ is determined from \eqref{eq:general_eikonal_character} and \eqref{eq:geo_general_ode}. \textcolor{black}{As a result, the probability currents defined by \eqref{eq:chiral_conserve_1} and \eqref{eq:anti_current} can be computed along the characteristic curves, where we find that the current direction coincides with the tangent to the curve}

As a simple illustration of the above, we consider a constant potential case, where $V=V_0$ is constant for both chiral and antichiral arrays. The eikonal $S$ is then of the form 
\begin{equation}
    S(x,y)=q_0x+p_0y+S_0 ,
\label{eq:phase_plane_wave}
\end{equation}
where $q_0$, $p_0$ and $S_0$ are constant.
It can be seen that the constants obey the relations
\begin{equation}
    (q_0+V_0)^2-p_0^2=1
\end{equation}
and 
\begin{equation}
    q_0^2+p_0^2=(V_0-1)^2 
\end{equation}
for the chiral and antichiral cases, respectively. Moreover, we find from \eqref{eq:chiral_geo_amplitude} and \eqref{eq:antichiral_geo_amplitude} that the amplitudes $u_0$ and $v_0$ are also constant. As may be expected, we conclude that a uniform potential gives rise to plane wave solutions. \textcolor{black}{The case of a piecewise constant potential is examined in Section. \ref{sec:scattering} C, where an analog of Snell's law emerges in the antichiral array, but is absent in the chiral array due to the prohibition of backscattering.}

Next, we consider a medium with a linear interface separating two half-planes. We assume that the potential $V$ is given by
\begin{equation}
V_{a,c}(x,y) =
\begin{cases}
V_0 \ , \quad &x < 0, \\
ax \ , \quad & x\ge 0 ,
\end{cases}
\end{equation} 
where $a$ is constant. Suppose that a unit amplitude plane wave is incident 
from the region $x<0$. We calculate the field by solving the eikonal and transport equations along the characteristic curves. Uniqueness of the solutions is assured by transmission conditions on the interface, which correspond to the continuity of the phase and the directions of the characteristic curves. We outline the main results here and present the detailed calculations in Appendix \ref{sec:app}.  In the chiral case, we find that the characteristic curves are straight lines, where the amplitudes $u_0$ and $v_0$ are constant along these lines. The only influence of the gradient comes from the phase $S$, which is given by
\begin{equation}
    S(x,y)=q_0x-\frac{a}{2}x^2+p_0y+S_0 ,
\label{eq:phase_linear_x}
\end{equation}
with $q_0$, $p_0$ and $S_0$ being constant.
Comparing \eqref{eq:phase_linear_x} with \eqref{eq:phase_plane_wave}, we note that an additional phase term emerges due to the linear gradient of the potential. 

\begin{figure}[t]
\centering
		\begin{tikzpicture}
			\node[inner sep=0pt] (OneReal) at (0,0) {\includegraphics[width=.22\textwidth]{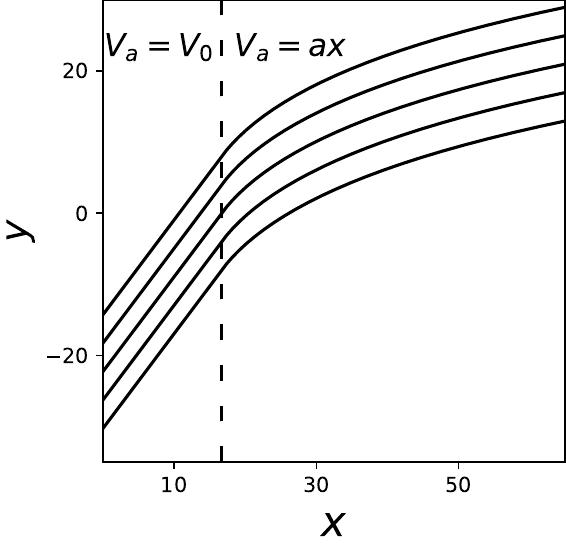}};
			\node[inner sep=0pt] (OneReal) at (4.5,0) {\includegraphics[width=.25\textwidth]{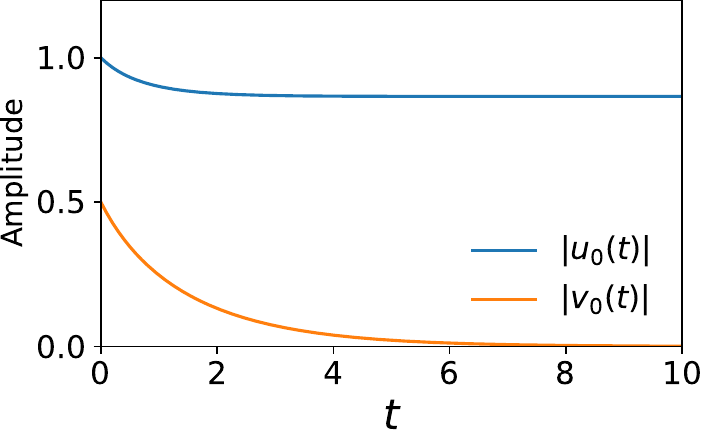}};
			\node at (-1.5,2) {(a)};
			\node at (3.1,1.8) {(b)};
		\end{tikzpicture}
\caption{Geometrical optics of the antichiral array with a constant potential gradient in the $x$ direction. The characteristic curves are shown in (a) and the field amplitudes in (b). {The parameters are chosen as $a=0.3$, $V_0 = 6$ and the incident transverse wavenumber is $p_0 = 4$.}}
\label{fig:geo_ax_anti}
\end{figure}

In the antichiral case, the characteristics are no longer straight lines in the region $x\ge 0$. They acquire a positive curvature and asymptotically become parallel to the gradient of the potential. We note that a similar phenomenon arises in geometrical optics of gradient index lenses~\cite{}. The characteristics are shown in FIG.~\ref{fig:geo_ax_anti} (a). Moreover, the amplitude $v_0$ decays to zero, as shown in FIG.~\ref{fig:geo_ax_anti} (b). 
This behavior is expected since in the antichiral case, the two subarrays are oriented in opposite directions. One subarray is aligned antiparallel to the gradient of the potential, preventing photon transport in this subarray. 

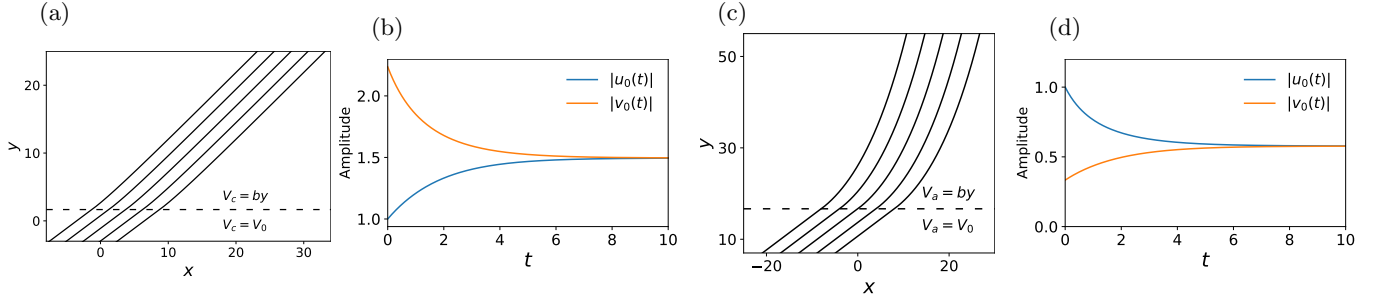
\begin{figure*}[t]
\centering
		\begin{tikzpicture}
			\node[inner sep=0pt] at (0,0) {\includegraphics[width=.24\textwidth]{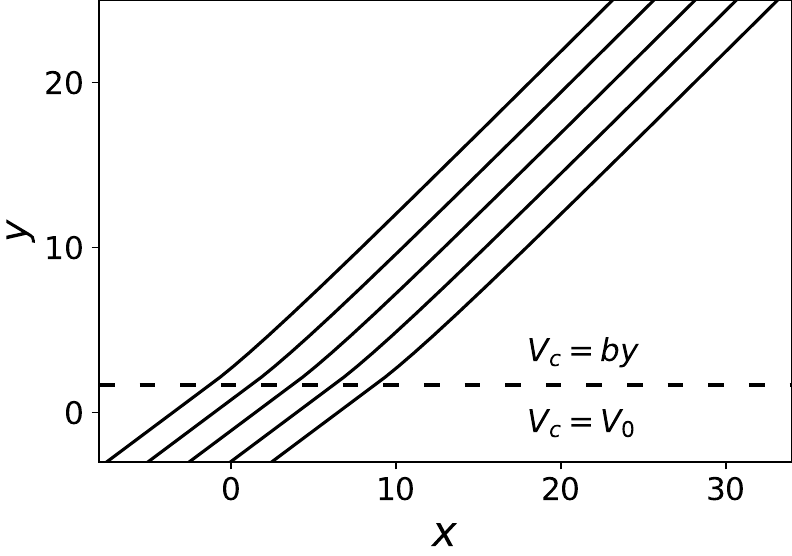}};
			\node[inner sep=0pt] at (4.5,0) {\includegraphics[width=.25\textwidth]{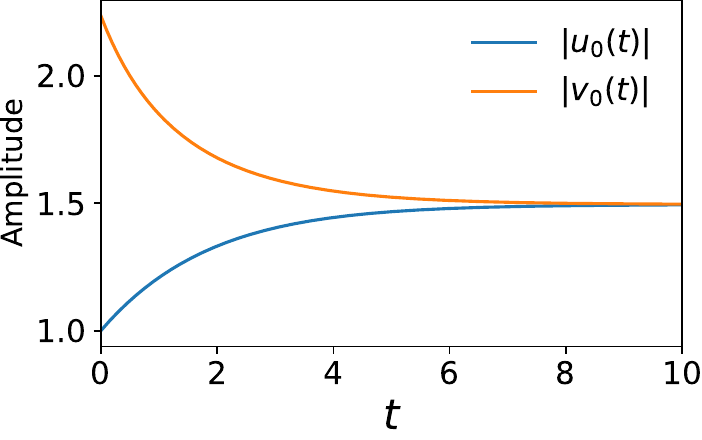}};
			\node at (-1.5,2) {(a)};
			\node at (2.9,1.8) {(b)};
    \begin{scope}[xshift=0.5\linewidth]
    
			\node[inner sep=0pt] at (0,0) {\includegraphics[width=.22\textwidth]{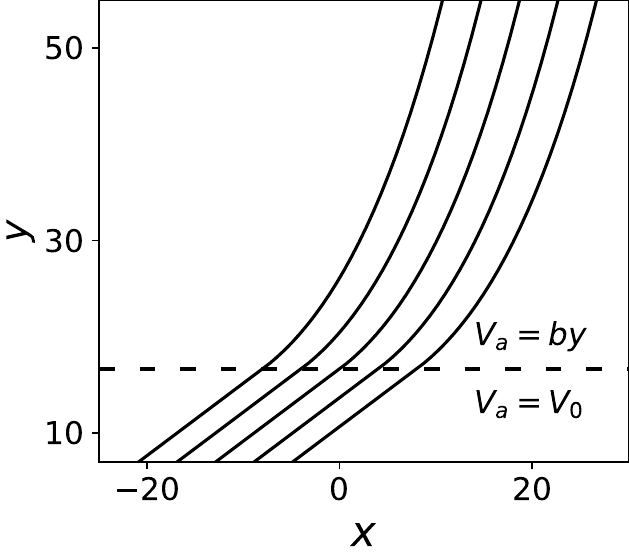}};
			\node[inner sep=0pt] at (4.5,0) {\includegraphics[width=.25\textwidth]{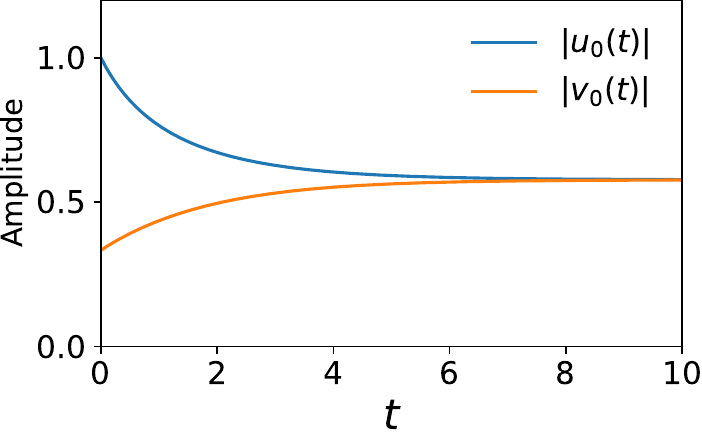}};
			\node at (-1.5,2) {(c)};
			\node at (2.9,1.8) {(d)};
    \end{scope}
		\end{tikzpicture}
     \caption{Solutions of the eikonal and transport equations for a constant potential gradient in the $y$ direction. We plot the characteristic curves in the chiral (a) and antichiral (c) arrays, and the corresponding field amplitudes in (b) and (d). The parameters are chosen as $b = 0.3$ and $V_0 = 0.5$ for the chiral array and $b = 0.3$, $V_0 = 6$ for the antichiral array. The incident wavenumber in the $x$ direction is taken to be $q_0 = 1$ and $4$ in the chiral and antichiral cases, respectively.}
\label{fig:geo_by}
\end{figure*}

Next, we assume that the gradient of the potential is in $y$ direction:
\begin{equation}
V_{a,c}(x,y) =
\begin{cases}
V_0 \ , \quad &y < 0, \\
by \ , \quad & y \ge 0 ,
\end{cases}
\end{equation} 
where $b$ is constant. We find that asymptotically, the characteristic curves deviate from the waveguide direction by a fixed angle of 45 degrees in the chiral case and 90 degrees in the antichiral case, as shown in FIG.~\ref{fig:geo_by} (a), (d). \textcolor{black}{In both arrays, the curves obtain positive curvatures as the potential gradient is applied along space-like directions.} The field amplitudes become asymptotically equal in two subarrays, \textcolor{black}{which reflects a geometric symmetry between the curves asymptotic direction and the waveguides alignment}, as shown in FIG.~\ref{fig:geo_by} (b), (d). 

\section{Diffraction}
\label{sec:diff}
In classical optics, diffraction refers to a scattering process in which the scatterer is
described by boundary conditions on the field rather than by a scattering potential. One paradigmatic example is light propagation through an aperture, where diffraction occurs when aperture size is comparable to the light wavelength. In this setting, geometrical optics fails to predict the emergent effects such as wave spreading and interference, which can be appropriately explained using diffraction theory. In this section, we examine analogous phenomena for waveguide arrays.

We begin by considering the boundary value problem for the Dirac equations  \eqref{eq:chiral_dirac_nondim} and \eqref{eq:anti_dirac_nondim} in the half-space $\{(x,y): x\ge0\}$. The field $\psi(0,y)$ is assumed to be specified on the line $x=0$ and the goal is to determine the field in the region $x>0$. We will solve this problem by deriving an integral representation for the field, which allows the boundary values at $x = 0$ to be propagated into the bulk region $x>0$.
 \textcolor{black}{We first consider the chiral case. We suppose that the potential $V$ vanishes for $x>0$ and denote by $\Tilde{\psi}(x,p)$ the Fourier transform of the field $\psi_c(x,y)$ with respect to $y$ according to
\begin{equation}
    \Tilde{\psi}_c(x,p)=\int_{-\infty}^{\infty}e^{-\mathrm{i}py} {\psi_c}(x,y) {dy}.\label{eq:diffrac_fourier_transform}
\end{equation}
It follows from \eqref{eq:chiral_dirac_nondim} that $\Tilde{\psi_c}(x,p)$ satisfies the equation
\begin{equation}
\mathrm{i}\partial_x \Tilde{\psi}_c + (p\alpha-k_c\beta)\Tilde{\psi}_c = 0 ,
\end{equation}
whose solution is given by
\begin{equation}
\Tilde{\psi}_c(x,p) =  \exp[\mathrm{i}(p\alpha - k_c\beta)x]\Tilde{\psi_c}(0,p) .
\label{fourier_chiral}
\end{equation}
Using the identity 
\begin{equation}
e^{\mathrm{i}\theta\hat{\mathbf{n}}\cdot\boldsymbol{\sigma}}
= \mathbb{1}\cos\theta + \mathrm{i}\hat{\mathbf{n}}\cdot\boldsymbol{\sigma}\sin\theta
\end{equation}
where $\boldsymbol{\sigma}=(\alpha,\beta)$ and inverting the Fourier transform, we find that \eqref{fourier_chiral} yields
\begin{equation}
    \psi_c(x,y)=\int_{-\infty}^{\infty}dy' K_c(x,y-y')\psi_c(0,y') .\label{eq:diffac_field_expr_progator}
\end{equation}
The propagator $K_c$ is defined by
\begin{equation}
    K_c(x,y)=\int_{-\infty}^{\infty}\frac{dp}{2\pi}e^{\mathrm{i}py}\biggl[\cos q_cx -ik_c \frac{\sin q_cx}{q_c}\beta + i p\frac{\sin q_cx}{q_c} \alpha\biggr].
    \label{eq: chiral_K_c_Green}
\end{equation}
Here $q_c=\sqrt{p^2+k_c^2}$ defines the chiral dispersion relation, which coincides with that of the one-dimensional Klein-Gordon equation. As a result, the propagator can be expressed in terms of the corresponding Green's function as
\begin{equation}
    K_c(x>0,y) = (-\partial_x+ik_c\beta-\alpha \partial_y)G_{KG}(\bm{r},\bm{0}),
\end{equation}
where 
\begin{equation}
G_{KG}(\bm{r},\bm{0}) = -\frac{1}{2}\Theta(x-|y|)J_0(k_c\sqrt{x^2-y^2}).
\end{equation}
}

 \textcolor{black}{Next, we consider the antichiral case. As before, we consider the Fourier transform of the field $\Tilde{\psi}_a(x,p)$. It follows from \eqref{eq:anti_dirac_nondim} that $\Tilde\psi_a(x,p)$ obeys
\begin{equation}
\mathrm{i}\beta\partial_{x}\Tilde\psi_a + (p\alpha -k_a\mathbb{1})\Tilde\psi_a=0 .
\label{eq:fourier_antichiral}
\end{equation}
\textcolor{black}{
A general solution to \eqref{eq:fourier_antichiral} is given by
\begin{equation}
    \Tilde{\psi}_{a}(x,p)=\begin{pmatrix}
        A_1^{+}\\
        A_2^{+}
    \end{pmatrix}e^{\mathrm{i}q_{a}x}+\begin{pmatrix}
        A_1^{-}\\
        A_2^{-}
    \end{pmatrix}e^{-\mathrm{i}q_{a}x},\label{eq:anti_fourier_sol}
\end{equation}
where $q_a=\sqrt{k_a^2-p^2}$ and the field amplitudes $A^{\pm}_{1,2}$ are to be determined. Noting that both coordinates $x$ and $y$ in the antichiral case are space-like, we apply the radiation condition to the field at infinity and set the inward-radiating amplitudes $A_{1,2}^{-} = 0$. The outward amplitudes $A_{1,2}^{+}$ can be determined by the boundary values $\Tilde{\psi}_a(0,y)$ in a similar manner to the chiral case. To this end, we find that the field $\psi_a$ takes the form
\begin{equation}
    \psi_a(x,y)=\int_{-\infty}^{\infty}dy' K_a(x,y-y')\psi_a(0,y'),
\label{eq:anti_diff_propa_exp}
\end{equation}
}
where the propagator $K_a$ is defined by
\begin{equation}
    K_a(x,y)=\int_{-\infty}^{\infty}\frac{dp}{2\pi}e^{\mathrm{i}py}e^{\mathrm{i}q_ax} = -2\partial_x G_H(\bm{r},\bm{0}).
\end{equation}
Here, $G_H(\bm{r},\bm{r}') = \frac{i}{4}H_0^{(1)}(k_a|\bm{r}-\bm{r}'|)$ is the Green's function of 2D Helmholtz equation, which admits the same dispersion relation as for $q_a$.
Eqs.~\eqref{eq:diffac_field_expr_progator} and \eqref{eq:anti_diff_propa_exp} can be used to propagate the fields $\psi_{c,a}$ from the boundary $x=0$ to any point in the half-space $x>0$. They are analogous to the first Rayleigh-Sommerfeld formula that arises in diffraction theory for classical light~\cite{carminati2021principles}.
}


We now illustrate the above results
with an example where the boundary values of the field are supposed to take the form
\begin{equation}
    \psi_{c,a}(0,y)=\Theta(a-|y|)\begin{pmatrix}
        0\\
        1
    \end{pmatrix}.\label{eq:diff_boundary_condition}
\end{equation}
The condition \eqref{eq:diff_boundary_condition} is an analogue of single-slit diffraction in classical optics. In our case, it can approximate the diffraction of a plane wave by a thin film placed at the origin with a slit of width $2a$, when the incident direction is perpendicular to the film. 

\begin{figure}[b]
\hspace{-0.5cm}
     \begin{subfigure}[b]{0.5\linewidth}
     \begin{tikzpicture}
			\node[inner sep=0pt] (0,0) {
         \includegraphics[width=1.2\textwidth]{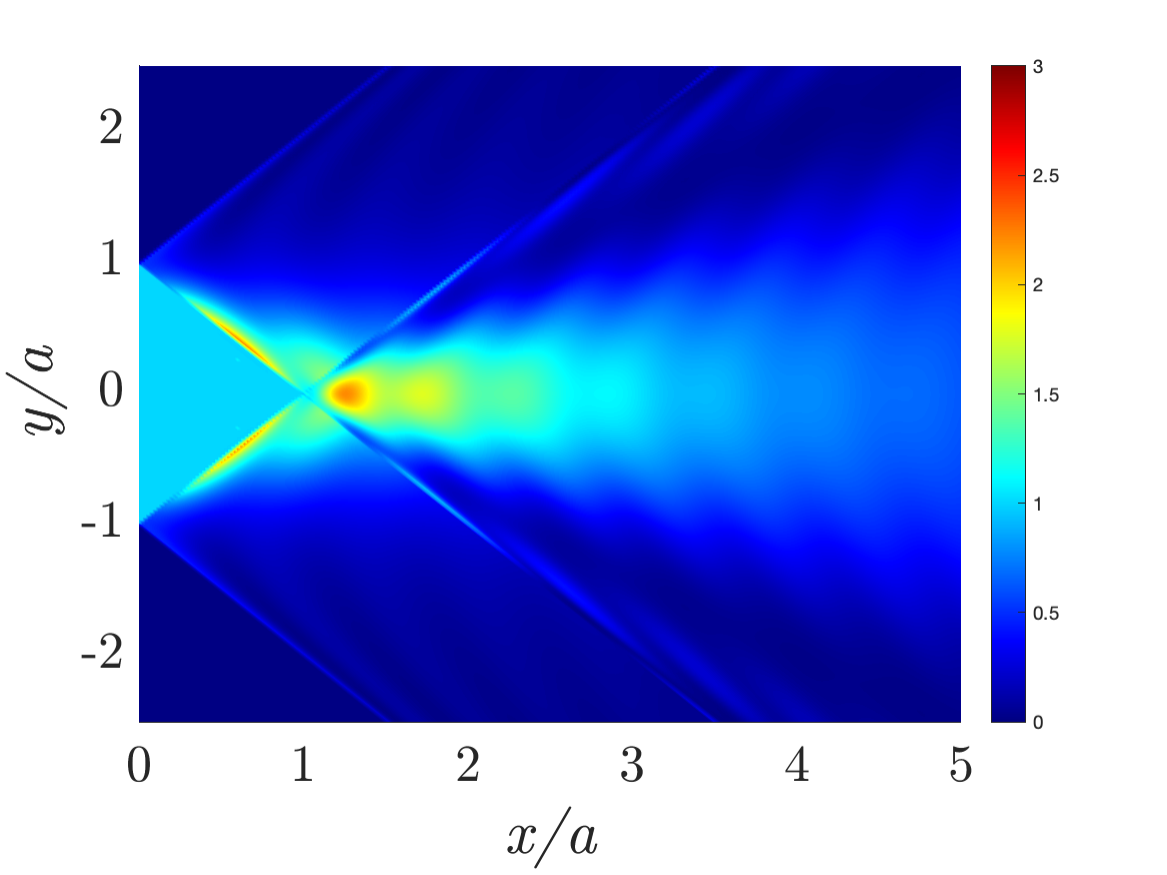}};
			\node at (-1.8,2) {(a)};
		\end{tikzpicture}
     \end{subfigure}\hspace{-0.1cm}
     \begin{subfigure}[b]{0.5\linewidth}
     \begin{tikzpicture}
			\node[inner sep=0pt] (0,0) {
         \includegraphics[width=1.2\textwidth]{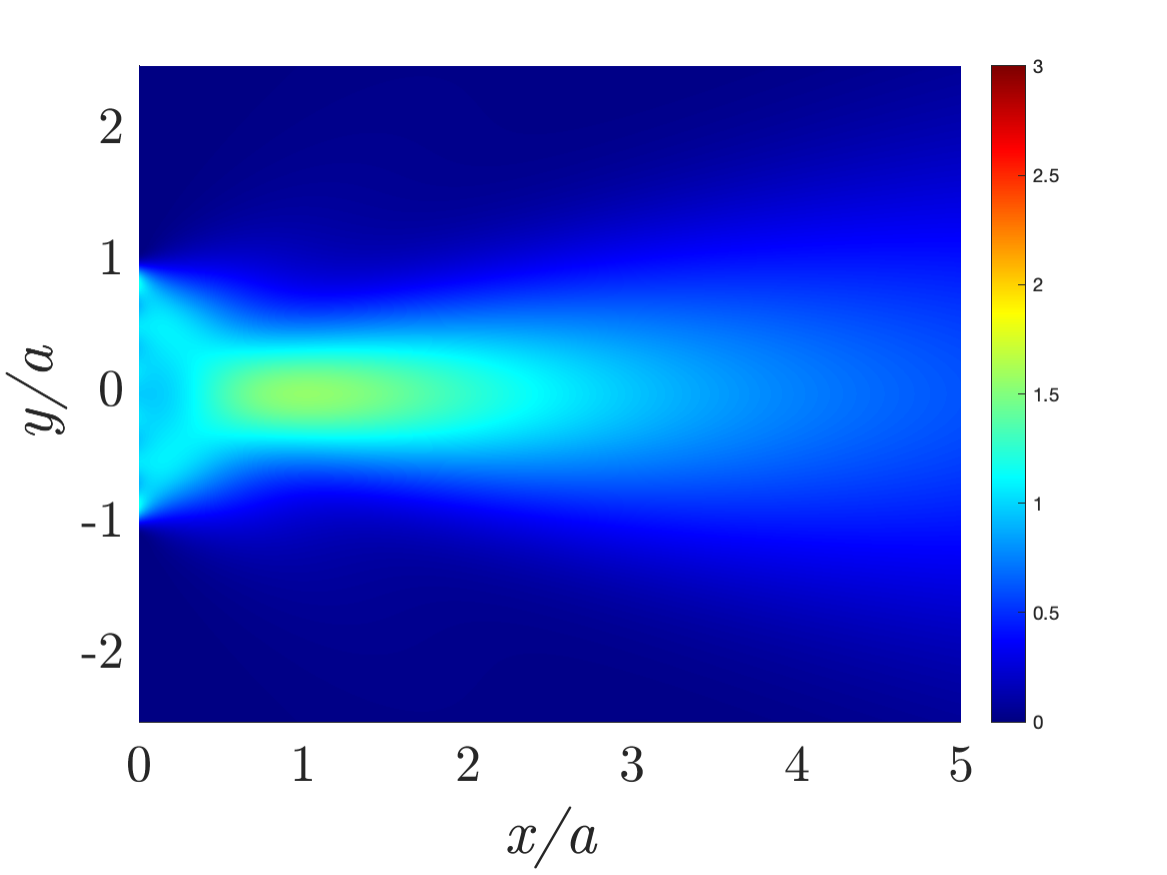}};
			\node at (-1.8,2) {(b)};
		\end{tikzpicture}
     \end{subfigure}
     \caption{The probability $|\psi|^2$ for a single-slit aperture in the chiral (a) and antichiral (b) arrays with $k_{a,c}a=5$.}
\label{fig:diff_chiral_anti}
\end{figure}

We begin with the chiral array. Inserting \eqref{eq: chiral_K_c_Green} and \eqref{eq:diff_boundary_condition} into \eqref{eq:diffac_field_expr_progator}, we obtain, for $x>0$,
\begin{equation}
\begin{split}
    \psi_{1c}(x,y)&=
        \int_{-a}^{a}dy'\,\biggl[-\frac{\text{sgn} (y-y')}{2}\delta\biggl(x-|y-y'|\biggr)J_0(k_c\xi')\\
        &\quad +\frac{k_c}{2}\Theta\biggl(x-|y-y'|\biggr)\frac{y-y'}{\xi'}J_1(k_c\xi')\biggr]\\
    \psi_{2c}(x,y)&=\int_{-a}^{a}dy'\,\biggl[\frac{1}{2}\delta\biggl(x-|y-y'|\biggr)J_0(k_c\xi')  \\
    &\quad+\frac{k_c}{2}\Theta\biggl(x-|y-y'|\biggr)\biggl(-\frac{x}{\xi'}J_1(k_c\xi')+ iJ_0(k_c\xi')\biggr)\biggr],\label{eq:chiral_diff_integral_slit}
\end{split}    
\end{equation}
where $\psi_{1c}$ and $\psi_{2c}$ denote the two components of the diffracted field and $\xi' = \sqrt{x^2 - |y-y'|^2}$. The 1\textsuperscript{st} component $\psi_{1c}$ can be evaluated in terms of Bessel functions as
\begin{multline}
    \psi_{1c}(x>0,y)= \frac{1}{2}\Theta\left(x^2-(y+a)^2\right)J_0\left(k_c\sqrt{x^2-(y+a)^2}\right)\\-\frac{1}{2}\Theta\left(x^2-(y-a)^2\right)J_0\left(k_c\sqrt{x^2-(y-a)^2}\right) \label{eq:chiral_1st_component_bessel}.
\end{multline}
The 2\textsuperscript{nd} component $\psi_{2c}$ in \eqref{eq:chiral_diff_integral_slit} cannot be expressed in terms of elementary functions. Instead, we examine its behavior in the limit $a\rightarrow0$, in which $\psi_{2c}$ is governed by
\begin{align}
    \psi_{2c}(a\rightarrow0)=&\frac{1}{2}\int_{-a}^{a}dy'\,\delta(x-|y-y'|)J_0(k_c\xi') \\
    =&\begin{cases}
1, &\text{if } |x-y|\leq a \text{ and } |x+y|\leq a,\\
\frac{1}{2}, &\text{if } |x-y|\leq a \text{ and } x+y\geq a, \\ & \qquad \text{ or } |x+y|\leq a \text{ and } x-y\geq a,\\
0, &\text{otherwise}.\label{eq:diff_chiral_2nd_component}
\end{cases}
\end{align}
We note from \eqref{eq:chiral_1st_component_bessel} and \eqref{eq:diff_chiral_2nd_component} that both field components vanish in the region $0<x<|y|-a$. In general, the diffraction pattern appears in the interior of \emph{light cones} $x=|y-y_0|$ for $y_0$ in the support of the boundary values \eqref{eq:diff_boundary_condition}. The formation of light cones is a unique feature of the chiral case. For the antichiral case, we note that the propagator $K_a$ is the same as the one for the 2D Helmholtz equation, indicating that the field diffraction will behave the same as well. Substituting \eqref{eq:diff_boundary_condition} into \eqref{eq:anti_diff_propa_exp}, we express the diffracted field as
\begin{equation}
    \psi_{a}(x>0,y)=\frac{ik_ax}{2}\int_{-a}^{a}dy'\frac{H_1^{(1)}\biggl(k_a\sqrt{x^2 + (y-y')^2}\biggr)}{\sqrt{x^2 + (y-y')^2}}\begin{pmatrix}
        0\\
        1
    \end{pmatrix}.
    \label{eq:diff_anti_integral_slit}
\end{equation}
We plot the 2D diffraction pattern according to $|\psi|^2=|\psi_{1}|^2+|\psi_{2}|^2$ in \Cref{fig:diff_chiral_anti} by numerically integrating \eqref{eq:chiral_diff_integral_slit} and \eqref{eq:diff_anti_integral_slit}.  We note that a clear light-cone feature appears in the chiral case but is absent in the antichiral case.

\section{Scattering}\label{sec:scattering}
We now study the case when potential $V$ in the Dirac equations \eqref{eq:chiral_dirac_nondim} and \eqref{eq:anti_dirac_nondim} has finite range, corresponding to a collection of scatterers. When illuminating the scatterers by an incident field $\psi_i$, the total field $\psi$ consists of incident and scattered parts and can be expressed as
\begin{equation}
    \psi=\psi_i+\psi_s.\label{eq:scatt_field_decompose}
\end{equation}
Here, the incident part $\psi_i$ obeys the Dirac equation with zero potential
\begin{equation}
    (-\mathrm{i}\partial_{x}+\mathrm{i}\alpha\partial_y+k_c\beta)\psi_i=0\label{eq:chiral_incident}
\end{equation}
in the chiral array and
\begin{equation}
(\mathrm{i}\beta\partial_x+\mathrm{i}\alpha\partial_y-k_a)\psi_i=0\label{eq:anti_inc_eq}
\end{equation}
in the antichiral array. It follows from \eqref{eq:chiral_incident} and \eqref{eq:anti_inc_eq} that the scattered field $\psi_s$ in the chiral and antichiral case satisfies
\begin{equation}
    (-\mathrm{i}\partial_x+D_c)\psi_s=-V_ck_c\psi \label{eq:scatt_chiral_eq}
\end{equation}
and
\begin{equation}
    (D_a-k_a)\psi_s=-V_ak_a\psi, \label{eq:scatt_anti_eq}
\end{equation}
where we define the Dirac operators according to
\begin{equation}
    D_c=\mathrm{i}\alpha\partial_y+k_c\beta,\quad D_a=\mathrm{i}\beta\partial_x+\mathrm{i}\alpha\partial_y.
\end{equation}
A key observation is that the operators acting on $\psi_s$ in \eqref{eq:scatt_chiral_eq} and \eqref{eq:scatt_anti_eq} can be inverted using the identities
\begin{align}
    (\mathrm{i}\partial_x+D_c)(-\mathrm{i}\partial_x+D_c) &=\partial^2_x-\partial^2_y+k_c^2,\\
    (D_a+k_a)(D_a-k_a) &=-(\Delta+k_a^2),\label{eq:operator_relation} 
\end{align}
which relate the problems to the Klein-Gordon and Helmholtz equations, respectively.
This enables a reformulation of equations \eqref{eq:scatt_chiral_eq} and \eqref{eq:scatt_anti_eq} as integral equations, which are given by
\begin{equation}
    \psi_s^{\text{chiral}}=(\mathrm{i}\partial_x+D_c)\int_S d^2r'G_{KG}(\bm{r},\bm{r}')V_c(\bm{r}')k_c\psi(\bm{r}')\label{eq:point_scatt_chiral_integral}
\end{equation}
and
\begin{equation}
    \psi_s^{\text{anti}}=-(D_a+k_a)\int_S d^2r' G_H(\bm{r},\bm{r}') V_a(\bm{r}')k_a\psi(\bm{r}')
    \label{eq:point_scatt_anti_integral}
\end{equation}
for the chiral and antichiral cases, respectively.
Here, we define $\bm{r} = (x,y)$ and use $G_{KG}(\bm{r},\bm{r}')$ and $G_{H}(\bm{r},\bm{r}')$ to denote the Green's functions of the 1D Klein-Gordon and 2D Helmholtz equations. We have
\begin{equation}
    G_{KG}(\bm{r},\bm{r}')=-\frac{1}{2}\Theta(x-x')\Theta(\xi^2)J_0(k_c\xi),
\end{equation}
and
\begin{equation}G_H(\bm{r},\bm{r}')=\frac{\mathrm{i}}{4}H_0^{(1)}(k_a|\bm{r}-\bm{r}'|),\end{equation}
where $\xi=\sqrt{(x-x')^2-(y-y')^2}$.

We proceed to derive an optical theorem for the Dirac equations. It follows from \eqref{eq:chiral_current} and \eqref{eq:anti_current} that the scattered probability current $\bm{j_s}$ can be defined by
\begin{equation}
	\bm{j_s}^\text{chiral}= \psi_s^{\dagger}\psi_s\hat{\bm{x}}-\psi_s^{\dagger}\alpha\psi_s \hat{\bm{y}},\quad
	\bm{j_s}^\text{anti}=-(\psi_s^{\dagger}\beta\psi_s\hat{\bm{x}}+\psi_s^{\dagger}\alpha\psi_s\hat{\bm{y}}),
	\label{eq:scattered_current}
\end{equation}
in the chiral and antichiral arrays, respectively.
The scattered probability $P_s$, which measures the total outward flux of the scattered current $\bm{j}_s$ across a boundary enclosing the scatterer $S$, can thereby be calculated as
\begin{equation}	P_s=\int_Sd^2r'\nabla\cdot{\bm{j_s}} = -2 k_{c,a}\operatorname{Im}\int_S d^2r' V_{c,a}\psi_i^{\dagger}\psi.
	\label{eq:scatt_prob_optic_theorem}
\end{equation}
Eq.~\eqref{eq:scatt_prob_optic_theorem} is an analogue of the optical theorem for the Dirac equations \eqref{eq:chiral_dirac_nondim} and \eqref{eq:anti_dirac_nondim} \cite{carminati2021principles}. Moreover, we can define the differential scattering cross section as
\begin{equation}
	\frac{d\sigma_s}{d\Omega}=\frac{dP_s/d\Omega}{|\bm{j_i}|}\label{eq:diff_cross_section},
\end{equation}
where $\bm{j_i}$ denotes the incident probability current. If $|\bm{j_i}|=1$, we obtain
\begin{equation}
	\sigma_s=\int d\Omega \frac{d\sigma_s}{d\Omega}=\int d\Omega \frac{dP_s}{d\Omega}=P_s\label{eq:cross_section}.
\end{equation} 
In the following, we solve the scattering problems in three configurations of interest. 

\subsection{Point scatterers} 
\begin{figure}[t]
\hspace{-0.5cm}
     \begin{subfigure}[b]{0.5\linewidth}
     \begin{tikzpicture}
			\node[inner sep=0pt] (0,0) {
         \includegraphics[width=1.2\textwidth]{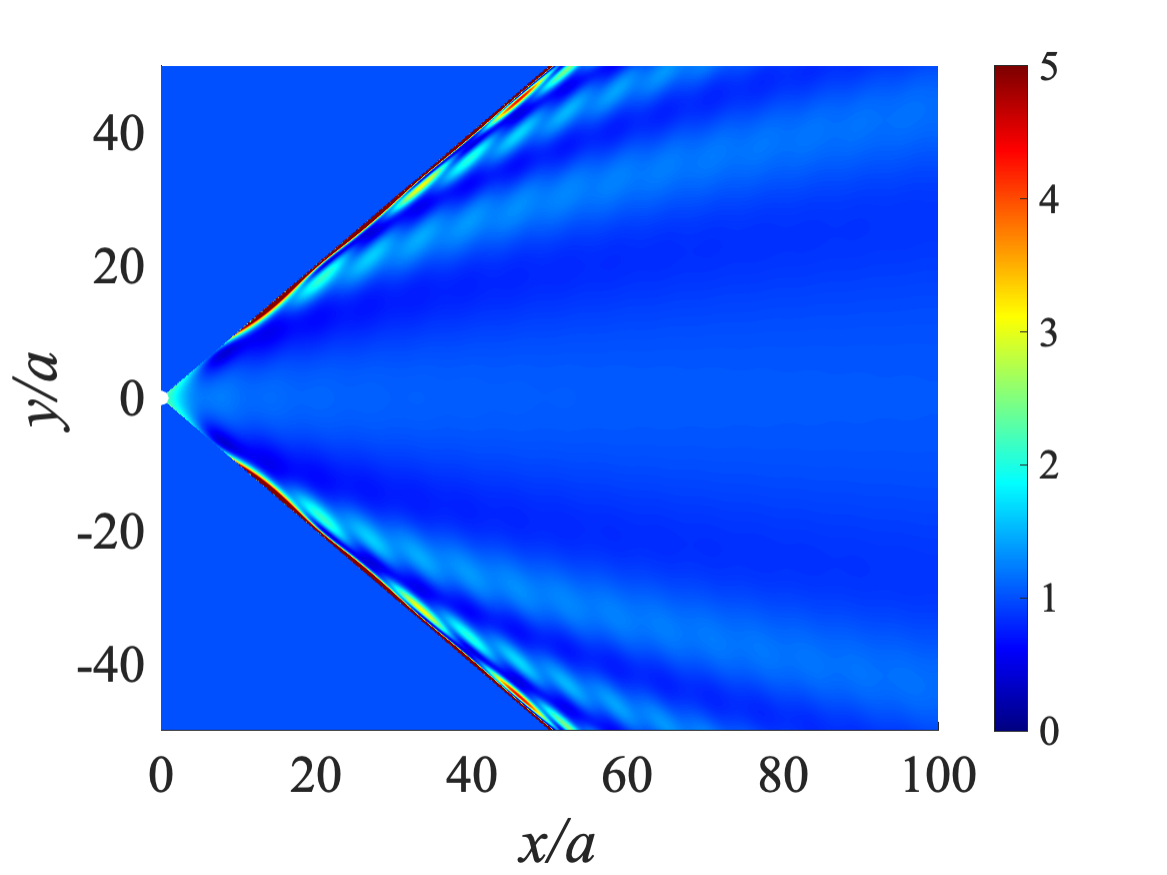}};
			\node at (-1.8,2) {(a)};
		\end{tikzpicture}
     \end{subfigure}\hspace{-0.1cm}
     \begin{subfigure}[b]{0.5\linewidth}
     \begin{tikzpicture}
			\node[inner sep=0pt] (0,0) {
         \includegraphics[width=1.2\textwidth]{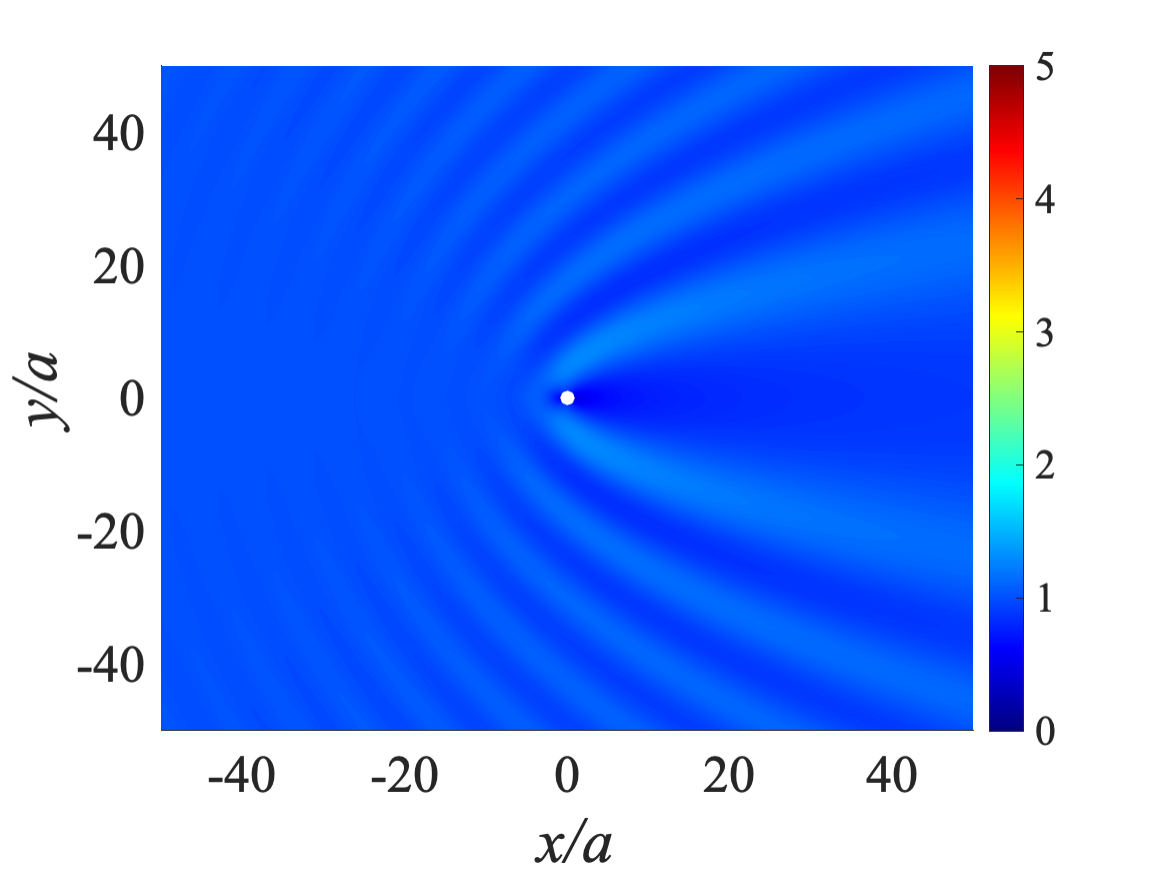}};
			\node at (-1.8,2) {(b)};
		\end{tikzpicture}
     \end{subfigure}
     \caption{The scattered probability $|\psi_s|^2$ of a single point particle in the chiral $(a)$ and antichiral $(b)$ array with $k_{c,a}a=0.5$ and $V_c=V_a=1.5$.}
\label{fig:single_scatt}
\end{figure}

In this subsection, we consider circular scatterers whose radius is much smaller than the wavelength of the incident field. We first examine a single scatterer centered at $\bm{r}=\bm{r}_0$ with radius $a$ and constant potential $V_{c,a}$. We assume that $k_{c,a}a \ll 1$, so that the scatterer can be treated as point-like and the field $\psi(\bm{r})$ is approximately uniform within its interior. Under this assumption, Eqs.~\eqref{eq:point_scatt_chiral_integral} and 
\eqref{eq:point_scatt_anti_integral} can be approximated as
\begin{equation}
    \psi_s^{\text{chiral}}(\bm{r})=V_ck_c\left((\mathrm{i}\partial_x+D_c)\int_S d^2r'G_{KG}(\bm{r},\bm{r}')\right)\psi(\bm{r}_0)\label{eq:point_scatt_chiral_integral_approx}
\end{equation}
and 
\begin{equation}
    \psi_s^{\text{anti}}(\bm{r})=-V_ak_a\left((D_a+k_a)\int_S d^2r' G_H(\bm{r},\bm{r}')\right) \psi(\bm{r}_0),
    \label{eq:point_scatt_anti_integral_approx}
\end{equation}
respectively.
We note that the scattered field at an observation point $\bm{r}$ is determined by the total field at the scatterer location $\bm{r}_0$. We proceed to compute $\psi(\bm{r}_0)$ by evaluating the above equations at $\bm{r}=\bm{r}_0$, which yields the self-consistent relations
\begin{equation}
(1+\lambda_c\beta+\mathrm{i}\gamma_c)\psi(\bm{r}_0)=\psi_i(\bm{r}_0) ,\quad (1+\lambda_a+\mathrm{i}\gamma_a)\psi(\bm{r}_0)=\psi_i(\bm{r}_0),\label{eq:single_tot_origin}
\end{equation}
where
\begin{align}
     \lambda_c&=V_c\pi(k_ca)^2/8, & \gamma_c&=V_ck_ca/\sqrt{2},\\ \lambda_a&=-V_a(k_aa)^2\log(k_aa)/2,& \gamma_a&=V_a\pi(k_aa)^2/4.
\end{align}
Once $\psi(\bm{r}_0)$ is determined from Eq.~\eqref{eq:single_tot_origin}, the scattered field can be obtained from \eqref{eq:point_scatt_chiral_integral_approx} and \eqref{eq:point_scatt_anti_integral_approx}, which gives
\begin{multline}
    \psi_s^{\text{chiral}}(\bm{r})=
    4\mathrm{i}\lambda_c\Theta(x-x_0)\Theta(\xi^2)\\ \times\left(\frac{(x-\alpha y)}{\xi}J_1(k_c\xi)+\mathrm{i}\beta J_0(k_c\xi)\right)(1+\lambda_c\beta+\mathrm{i}\gamma_c)^{-1}\psi_i(\bm{r}_0)
    \label{eq:scatt_chiral_single_expr}
\end{multline}
and 
\begin{multline}
    \psi_s^{\text{anti}}(\bm{r})=-\frac{\gamma_a}{1+\lambda_a+\mathrm{i}\gamma_a}\bigg(\beta H_1^{(1)}(k_a|\bm{r}|)\frac{x}{|\bm{r}|}\\ +\alpha H_1^{(1)}(k_a|\bm{r}|)\frac{y}{|\bm{r}|}+\mathrm{i}H_0^{(1)}(k_a|\bm{r}|)\bigg)\psi_i(\bm{r}_0),\label{eq:scatt_anti_single_expr}
\end{multline}
for the chiral and antichiral arrays. Here, we have assumed that the observation point $\bm{r}$ is far away from the point-like scatterer and approximate
\begin{align}
    \int_S d^2r' G(\bm{\rho},\bm{r}')&\approx S_0G(\bm{r},\bm{r}_0),\label{eq:scatt_Green_integral_approx}
\end{align}
where $G = G_{KG}$ or $G_H$ denotes the Green's function and $S_0$ is the area of the scatterer. Upon inserting  \eqref{eq:scatt_chiral_single_expr} and \eqref{eq:scatt_anti_single_expr} into \eqref{eq:cross_section}, we obtain the cross sections as
\begin{equation}
\begin{aligned}
    \sigma_s^{\text{chiral}}/2a&=\frac{\pi V_c^2(k_ca)^2}{\sqrt{2}
    \bigl((1-\lambda_c)^2+\gamma_c^2\bigr)},\\ \sigma_s^{\text{anti}}/2a&=\frac{\pi^2V_a^2(k_aa)^3}{4\bigl((1+\lambda_a)^2+\gamma_a^2\bigr)}.
\label{eq:point_scatt_cross_section}
\end{aligned}
\end{equation}
\Cref{fig:single_scatt} shows the scattered field from a single point scatterer in both the chiral and antichiral arrays. The incident field is chosen as a plane wave with unit amplitude, propagating in positive $x$ direction. We note that the chiral scattering exhibits a light-cone feature whereas the antichiral scattering resembles that of classical optics.

\begin{figure}[t]
\hspace{-0.5cm}
     \begin{subfigure}[b]{0.5\linewidth}
     \begin{tikzpicture}
			\node[inner sep=0pt] (0,0) {
         \includegraphics[width=1.2\textwidth]{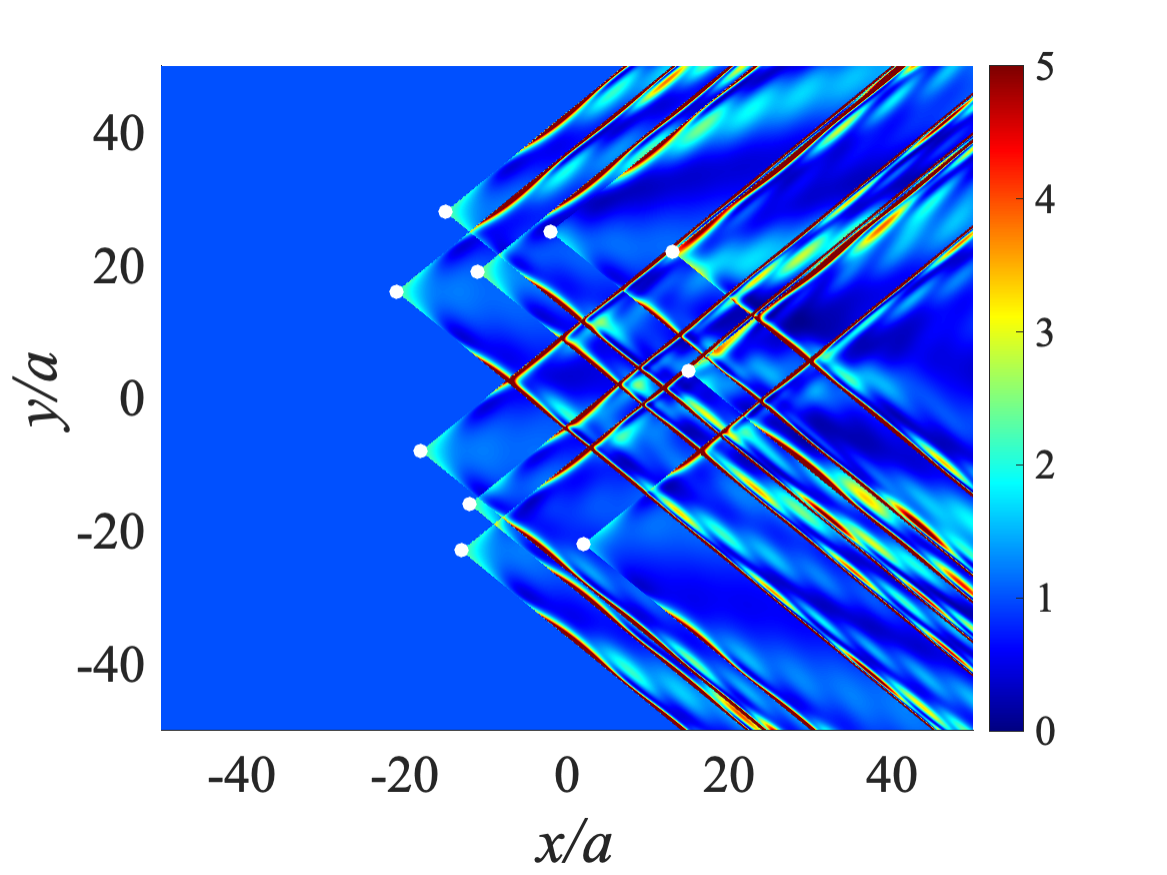}};
			\node at (-1.8,2) {(a)};
		\end{tikzpicture}
     \end{subfigure}\hspace{-0.1cm}
     \begin{subfigure}[b]{0.5\linewidth}
     \begin{tikzpicture}
			\node[inner sep=0pt] (0,0) {
         \includegraphics[width=1.2\textwidth]{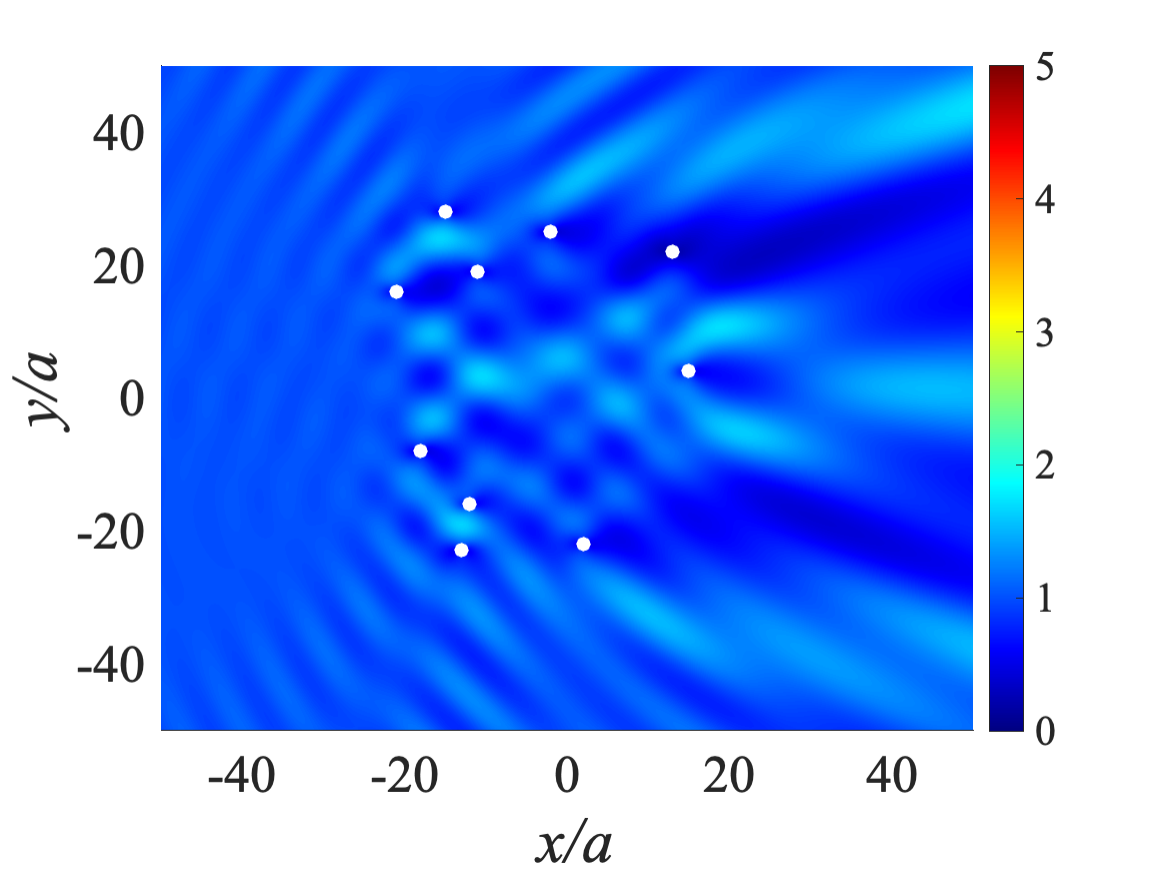}};
			\node at (-1.8,2) {(b)};
		\end{tikzpicture}
     \end{subfigure}
     \caption{The scattered probability $|\psi|^2$ of $N=10$ randomly distributed point particles in the chiral (a) and antichiral (b) array with $k_{c,a}a=0.5$ and $V_c=V_a=1.5$. The particles are labeled as white dots.}
\label{fig:multi_10_scatt}
\end{figure}

The above point scattering regime can be readily extended to a collection of point scatterers. As an illustration, we consider $N$ point scatterers at the positions $\bm{r}=\bm{r}_j$ for $j=1,2,\cdots N$ and suppose that the potential $V_{c,a}$ is constant inside the scatterers and vanishes outside. As a result, Eqs.~\eqref{eq:point_scatt_chiral_integral_approx} and \eqref{eq:point_scatt_anti_integral_approx} are now modified as
\begin{equation}
    \psi_s^\text{chiral}(\bm{r})=V_ck_c\sum_{j=1}^{N}\left((\mathrm{i}\partial_x+D_C)\int_{S_j} d^2r'G_{KG}(\bm{r},\bm{r}')\right)\psi(\bm{r}_j)\label{eq:chiral_multi_scatt_int},
\end{equation}
and
\begin{equation}
   \psi_s^{\text{anti}}(\bm{r})=-V_ak_a\sum_{j=1}^N \left((D_A+k_a)\int_{S_j} d^2r' G_H(\bm{r},\bm{r}')\right) \psi(\bm{r}_j)\label{eq:anti_multi_scatt_int},
\end{equation}
respectively, where the summation is taken over all scatterers. Upon evaluating Eqs.~\eqref{eq:chiral_multi_scatt_int} and \eqref{eq:anti_multi_scatt_int} at $\bm{r} = \bm{r}_0$, we obtain 
\begin{equation}
S_{c,a}\left(
\begin{smallmatrix}
    \psi(\bm{r}_1)\\
    \svdots\\
    \psi(\bm{r}_j)\\
    \svdots\\
    \psi(\bm{r}_N)
\end{smallmatrix}\right)
=\left(\begin{smallmatrix}
    \psi_i(\bm{r}_1)\\
    \svdots\\
    \psi_i(\bm{r}_j)\\
    \svdots\\
    \psi_i(\bm{r}_N)
\end{smallmatrix}\right),
\label{eq:chiral_multi_matrix_tot_inc}
\end{equation}
where
\begin{equation}
\begin{split}
    S_c&=-4\mathrm{i}\lambda_c
    \left(\begin{smallmatrix}
        &(1+\lambda_c\beta+\mathrm{i}\gamma_c)/4\mathrm{i}\lambda_c &\cdots &h_c(\bm{r}_1,\bm{r}_N)\\
        &\svdots &\sddots &\svdots\\
        &h_c(\bm{r}_N,\bm{r}_1)&\cdots &(1+\lambda_c\beta+\mathrm{i}\gamma_c)/4\mathrm{i}\lambda_c
    \end{smallmatrix}\right)
\end{split}\label{eq:chiral_multi_scatt_matrix}
\end{equation}
and
\begin{equation}
    S_a=
    \left(\begin{smallmatrix}
        &1+\lambda_a+\mathrm{i}\gamma_a &\cdots&h_a(\bm{r}_1,\bm{r}_N)\\
        &\svdots &\sddots &\svdots\\
        &h_a(\bm{r}_N,\bm{r}_1)&\cdots &1+\lambda_a+\mathrm{i}\gamma_a
    \end{smallmatrix}\right),
\end{equation}
are the scattering matrices and 
\begin{equation}
\begin{split}
    h_c(\bm{r}_j,\bm{r}_l)&=-4\mathrm{i}\lambda_c\bigl((x_j-\alpha y_j)J_1(k_c\xi_{jl})/\xi_{jl}+\mathrm{i}\beta J_0(k_c\xi_{jl})\bigr),\\
    h_a(\bm{r}_j,\bm{r}_l)&=-\mathrm{i}(\beta x_j+\alpha y_j)\frac{H_1^{(1)}(k_a|\bm{r}_j-\bm{r}_l|)}{|\bm{r}_j-\bm{r}_l|}+H_0^{(1)}(k_a|\bm{r}_j-\bm{r}_l|),\label{eq:anti_multi_h}
\end{split}
\end{equation}
are auxiliary functions derived from the approximation \eqref{eq:scatt_Green_integral_approx}. In addition, $\xi_{jl}=\sqrt{(x_j-x_l)^2-(y_j-y_l)^2}$. Inserting \eqref{eq:chiral_multi_matrix_tot_inc} into \eqref{eq:chiral_multi_scatt_int} and \eqref{eq:anti_multi_scatt_int} and using \eqref{eq:scatt_Green_integral_approx} to simplify the resulting equations, we can express the scattered field in terms of the incident field according to 
\begin{equation}
    \psi_s(\bm{r})=-
    \left(\begin{smallmatrix}
        &h_{c,a}(\bm{r},\bm{r}_1) &0 &\cdots &0\\
        &0 &h_{c,a}(\bm{r},\bm{r}_2)&\cdots &0\\
        &\svdots &\svdots &\sddots &\svdots\\
        &0&0&\cdots&h_{c,a}(\bm{r},\bm{r}_N)
    \end{smallmatrix}\right)
    S_{c,a}^{-1}
    \left(\begin{smallmatrix}
    \psi_i(\bm{r}_1)\\
    \psi_i(\bm{r}_2)\\
    \svdots\\
    \psi_i(\bm{r}_N)
    \end{smallmatrix}\right).\label{eq:scatt_chiral_multiple_field_expr_inc}
\end{equation}
    In \Cref{fig:multi_10_scatt}, we plot $|\psi|^2=|\psi_i+\psi_s|^2$ based on \eqref{eq:scatt_chiral_multiple_field_expr_inc} with $N=10$ point scatterers distributed randomly. We observe that each scatterer in the chiral array gives rise to a  light cone, within which interference between scattered field is limited, whereas the antichiral scattering exhibits smooth interference effects. These distinctions underscore the fundamental difference in photon transport between the two types of waveguide arrays.

\subsection{Mie scattering in antichiral arrays}
\begin{figure}[t]
\hspace{-0.7cm}
     \begin{subfigure}[b]{0.5\linewidth}
     \begin{tikzpicture}
			\node[inner sep=0pt] (0,0) {
         \includegraphics[width=1.3\textwidth]{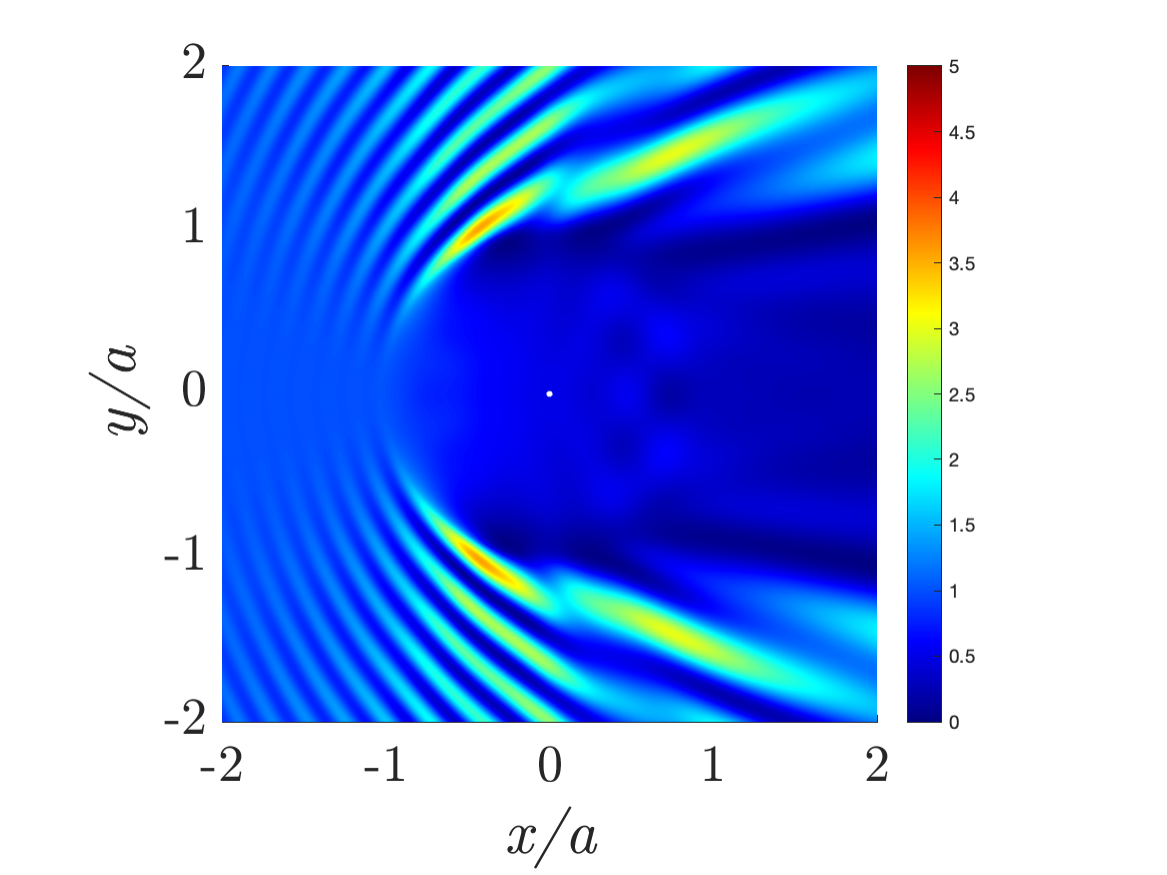}};
			\node at (-1.6,2) {(a)};
		\end{tikzpicture}
     \end{subfigure}
\hspace{-0.2cm}
     \begin{subfigure}[b]{0.5\linewidth}
     \begin{tikzpicture}
			\node[inner sep=0pt] (0,0) {
         \includegraphics[width=1.3\textwidth]{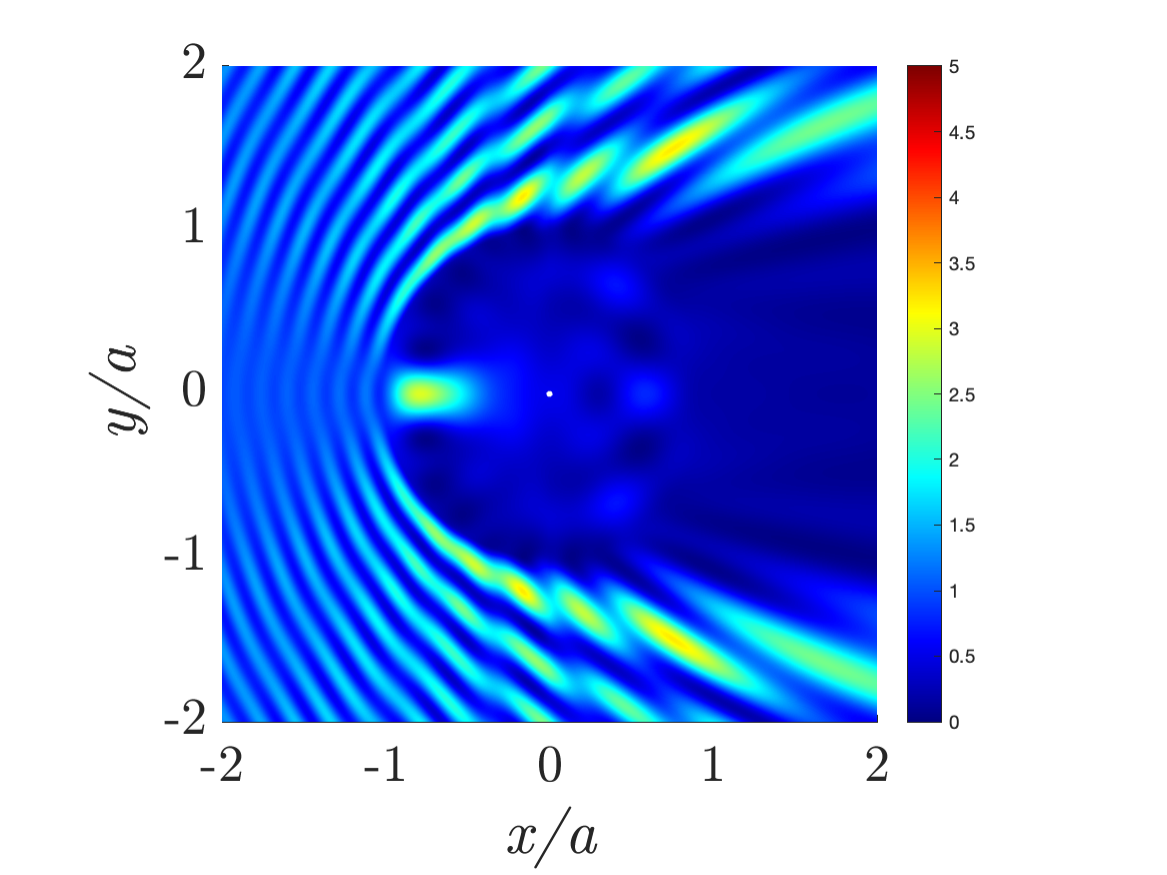}};
			\node at (-1.6,2) {(b)};
		\end{tikzpicture}
     \end{subfigure}
\caption{The probability $|\psi|^2$ from scattering by a circular particle  in the antichiral array with incident field given by \eqref{eq:anti_mie_inc} and  with $k_a a=20$ and the index (a) $V_a=0.5$ and (b) $V_a=1.5$. In both cases, we compute the scattered field using Eqs.~\eqref{eq:anti_mie_boudary_coeff}, \eqref{eq: mie_AB_expr} and \eqref{eq:mie_f_expr}.}
\label{fig:anti_mie_scatt_patt}
\end{figure}

In this subsection, we consider the scattering of a plane wave from a circular scatterer, which serves as the analogue of 2D Mie scattering in classical optics. We restrict our discussion to the antichiral case, since the chiral Dirac operator does not admit rotational symmetry. We begin with Eq.~\eqref{eq:anti_dirac_nondim}, which can be reformulated as a 2D Helmholtz equation:
\begin{equation}
\Bigl(\partial_x^2+\partial_y^2+\eta^2(\bm{r})k_a^2\Bigr)\psi=0\label{eq:anti_helmholtz}
\end{equation}
by multiplying the operator $D_a-(V_a-1)k_a$ on both sides.
Here, we define $\eta(\bm{r})=1-V_a(\bm{r})$ and suppose that the potential $V_a(\bm{r})$ is given by
\begin{equation}
V_a(\bm{r})=
\begin{cases}
V_0, \quad &|\bm{r}|\leq a,\\
0, \quad &|\bm{r}|> a.
\end{cases}
\label{eq:mie_potential_distribution}
\end{equation}
Moreover, we assume that the incident wave propagates along the waveguide, and takes the plane wave form according to
\begin{equation}
    \psi_{i}=\Bigl(
\begin{smallmatrix}
    0\\
    1
\end{smallmatrix}\Big)
e^{ik_ax}\label{eq:anti_mie_inc}.
\end{equation}
We note that both the Dirac equation \eqref{eq:anti_dirac_nondim} and the incident wave \eqref{eq:anti_mie_inc} is invariant under inversion $y\rightarrow -y$ if we simultaneously transform the field components according to $-\psi_{1}(x,-y)\rightarrow\psi_{1}(x,y)$ and $\psi_{2}(x,-y)\rightarrow\psi_{2}(x,y)$. This means that the two field components have distinct parities with respect to the $y$-axis, which yields
\begin{equation}
        \psi_1(x,y)=-\psi_1(x,-y),\quad
        \psi_2(x,y)=\psi_2(x,-y).
    \label{eq:anti_mie_symmetry}
\end{equation}
As a result, the total field outside and inside the circular scatterer can be expanded in terms of the cylindrical harmonics as
\begin{equation}
\begin{cases}
\psi_1^{\text{out}}(\bm{r})=\sum\limits_{n=1}^{\infty}A_nH_n^{(1)}(k_a|\bm{r}|)\sin n\theta,\\
    \psi_2^{\text{out}}(\bm{r})=e^{\mathrm{i}k_ax}+\sum\limits_{n=0}^{\infty}B_nH_n^{(1)}(k_a|\bm{r}|)\cos n\theta,
\end{cases}
\label{eq:anti_mie_out_expansion}
\end{equation}
and 
\begin{equation}
    \begin{cases}
\psi_1^{\text{in}}(\bm{r})=\sum\limits_{n=1}^{\infty}C_nJ_n(|\mu| k_a|\bm{r}|)\sin n\theta,\\
        \psi_2^{\text{in}}(\bm{r})=\sum\limits_{n=0}^{\infty}D_nJ_n(|\mu| k_a|\bm{r}|)\cos n\theta,
    \end{cases}
    \label{eq:anti_mie_in_expansion}
\end{equation}
 where $\mu= 1 - V_0$ and $\theta$ is the angle measured from the positive $x$-axis. In addition, the Hankel functions $H_n^{(1)}$ are used to satisfy the outgoing radiation condition as $\bm{r} \to \infty$. We proceed to determine the unknown coefficients $\{A_n, B_n, C_n, D_n\}$ as follows. Inserting \eqref{eq:anti_mie_out_expansion} and \eqref{eq:anti_mie_in_expansion} into \eqref{eq:anti_dirac_nondim} yields
\begin{equation}
    \begin{cases}
    A_{n+1}-B_{n+1}=-2\mathrm{i}A_n+(A_{n-1}+B_{n-1}),&n\geq2,\\
    A_2-B_2=-2\mathrm{i}A_1+2B_0, &n=1,\\
    A_{n+1}-B_{n+1}=-2\mathrm{i}B_n-(A_{n-1}+B_{n-1}),&n\geq2,\\
    A_2-B_2=-2\mathrm{i}B_1-2B_0,&n=1,\\
    A_1-B_1=-2\mathrm{i}B_0,&n=0,
    \end{cases}
    \label{eq:anti_mie_outside_1st_relation}
\end{equation}
outside the scatterer and
\begin{equation}
    \begin{cases}
    C_{n+1}-D_{n+1}=-2\mathrm{i}\mu/|\mu| C_n+(C_{n-1}+D_{n-1}), &n\geq2,\\
    C_2-D_2=-2\mathrm{i}\mu/|\mu| C_1+2D_0, &n=1,\\
    C_{n+1}-D_{n+1}=-2\mathrm{i}\mu/|\mu| D_n-(C_{n-1}+D_{n-1}), &n\geq2,\\
    C_2-D_2=-2\mathrm{i}\mu/|\mu| D_1-2D_0, &n=1,\\
    C_1-D_1=-2\mathrm{i}\mu/|\mu| D_0, &n=0,
    \end{cases}
    \label{eq:anti_mie_inside_1st_relation}
\end{equation}
inside the scatterer. A matching condition is applied to the boundary $|\bm{r}|=a$ to ensure the continuity of the field, which gives
\begin{equation}
    \begin{cases}
        C_n=H_n^{(1)}(k_aa)/J_n(|\mu| k_aa)A_n, &n\geq1,\\
        D_n=H_n^{(1)}(k_aa)/J_n(|\mu| k_aa)B_n+2\mathrm{i}^n J_n(k_aa)/J_n(|\mu| k_aa), \quad  & n\geq1,\\
        D_0=H_0^{(1)}(k_aa)/J_0(|\mu| k_aa)B_0+ J_0(k_aa)/J_0(|\mu| k_aa), &n=0.
    \end{cases}
    \label{eq:anti_mie_boudary_coeff}
\end{equation}
Solving Eqs.~\eqref{eq:anti_mie_outside_1st_relation}, \eqref{eq:anti_mie_inside_1st_relation} and \eqref{eq:anti_mie_boudary_coeff}, we obtain
\begin{equation}
\begin{cases}
    A_n=
    (f_n-\mathrm{i}f_{n-1})/2,\quad & n \geq 1,\\
    B_n=(f_n+\mathrm{i}f_{n-1})/2, \quad &n\geq 1,\\
    B_0=f_{0}/2,
\end{cases}
\label{eq: mie_AB_expr}
\end{equation}
where, for $n\geq0$,
\begin{equation}
    f_n=2\mathrm{i}^n\frac{-\mu/|\mu| J_{n+1}(k_aa)J_n{(|\mu| k_aa)}+J_n(k_aa)J_{n+1}(|\mu| k_aa)}{H_{n+1}^{(1)}(k_aa)J_n(|\mu|k_aa)-H_n^{(1)}(k_aa)J_{n+1}(|\mu|k_aa)}.
    \label{eq:mie_f_expr}
\end{equation}
Once $\{A_n, B_n\}$ are determined, the coefficients $\{C_n, D_n\}$ can be calculated according to \eqref{eq:anti_mie_boudary_coeff}. We plot the total field $|\psi|^2$ with $k_aa=20$ in \Cref{fig:anti_mie_scatt_patt} for two choices of the potential $V_a$ (0.5 and 1.5, respectively). We note that the two cases obey the same Helmholtz equations but different Dirac equations.
We next calculate the cross section given by \eqref{eq:cross_section}. Upon substituting \eqref{eq:anti_mie_in_expansion} into \eqref{eq:scatt_prob_optic_theorem}, we obtain
\begin{equation}
    \frac{\sigma_s}{2a}=-\frac{2\pi V_a}{\delta^2-1}\operatorname{Im}\sum_{n=0}^{\infty}(-\mathrm{i})^n D_n\zeta_n\label{eq:anti_mie_cross_series},
\end{equation}
where $\zeta_n$ is defined by
\begin{equation}
    \zeta_n=J_{n-1}(k_aa)J_n(|\mu| k_aa)-|\mu| J_{n-1}(|\mu| k_aa)J_n(k_aa). 
\end{equation}

\begin{figure}[b]
     \begin{subfigure}[b]{0.5\linewidth}
     \begin{tikzpicture}
			\node[inner sep=0pt] (0,0) {
         \includegraphics[width=1\textwidth]{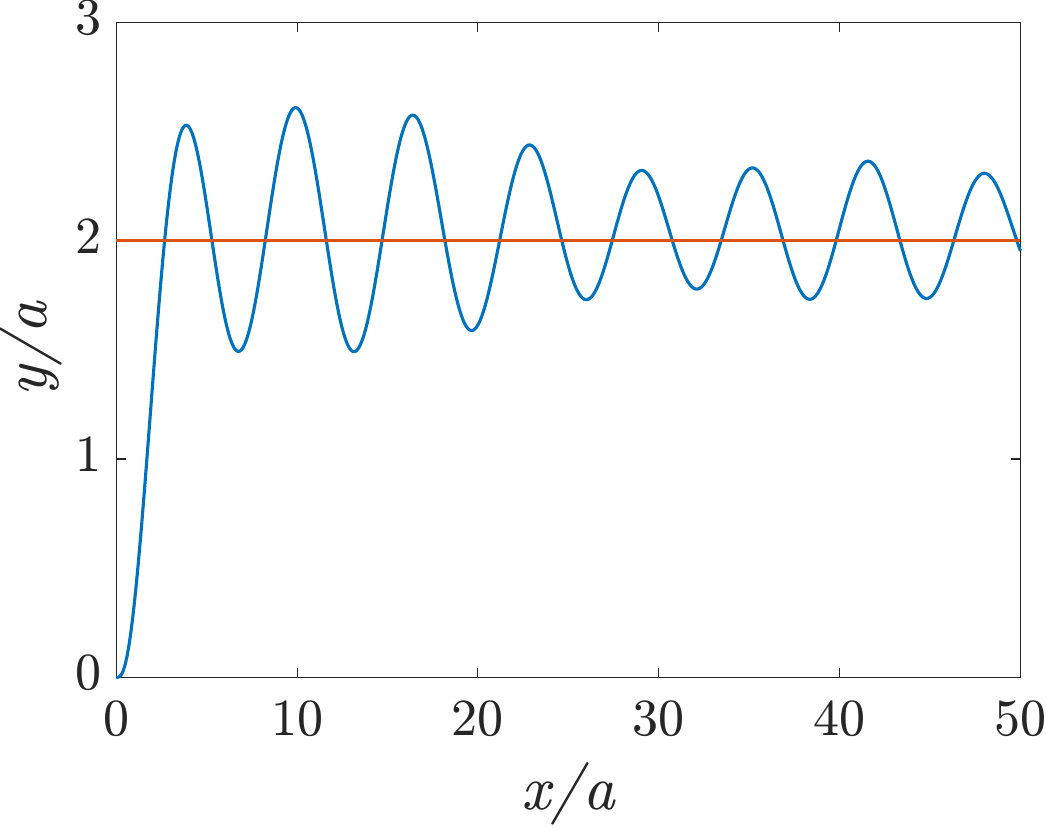}};
			\node at (-1.6,2) {(a)};
		\end{tikzpicture}
     \end{subfigure}
     \begin{subfigure}[b]{0.5\linewidth}
     \begin{tikzpicture}
			\node[inner sep=0pt] (0,0) {
         \includegraphics[width=1\textwidth]{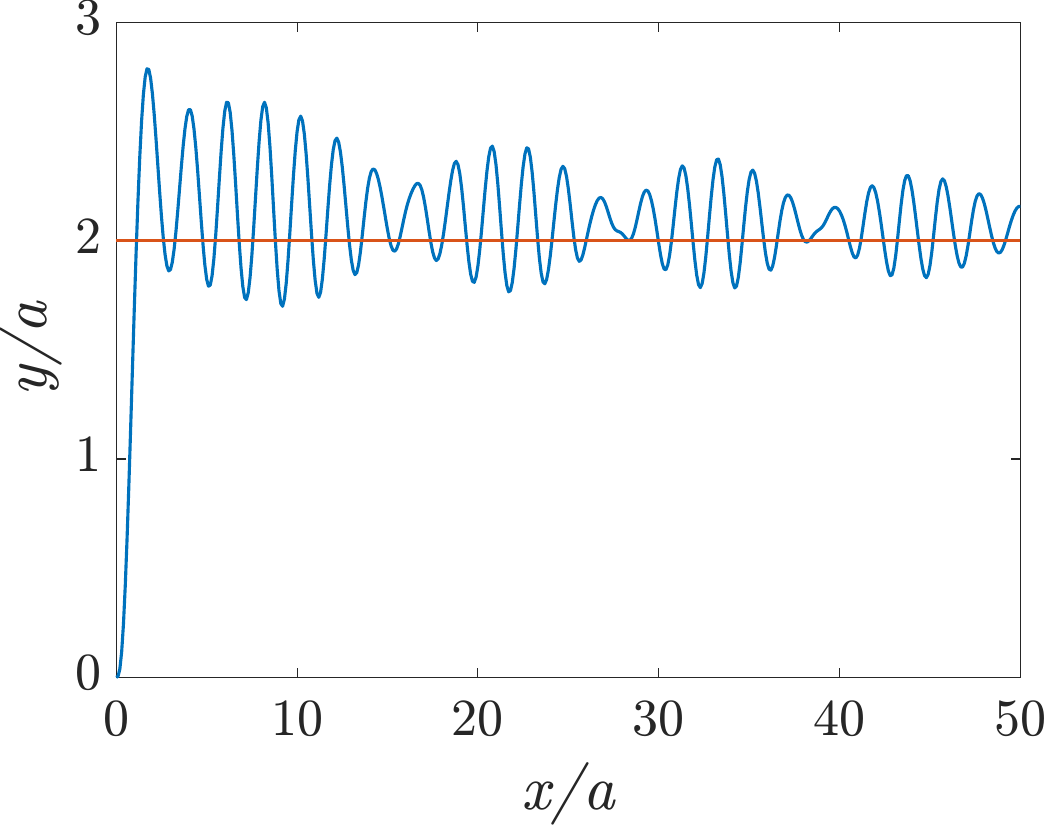}};
			\node at (-1.6,2) {(b)};
		\end{tikzpicture}
     \end{subfigure}
\caption{Frequency dependence of the normalized scattering cross section $\sigma_s/2a$ in the antichiral waveguide array with the index (a) $
V_a=0.5$ and (b) $
V_a=1.5$. In both cases, the scattering cross section approaches the value $2$ for large $k_aa$.}
\label{fig:anti_mie_section_plot}
\end{figure}

We plot the cross section $\sigma_s/2a$ as a function of $k_aa$ in \Cref{fig:anti_mie_section_plot}.
In the limit $k_aa\ll1$, \eqref{eq:anti_mie_cross_series} is dominated by its lowest order term, leading to an asymptotic form of the cross section as
\begin{equation}
\frac{\sigma_s}{2a}=\frac{V_a^2\pi^2}{4}(k_aa)^3.\label{eq:anti_mie_cross_section_limit}
\end{equation}
We note that \eqref{eq:anti_mie_cross_section_limit} is consistent with the result \eqref{eq:point_scatt_cross_section} obtained from the point scattering approximation. In contrast, in the limit $k_aa\to \infty$, the cross section approaches the limiting value $\sigma_s/2a = 2$. This is an analogue of the extinction paradox in classical optics \cite{carminati2021principles}.
                         
\subsection{Scattering from a slab}\label{sec: slab_scatter}
Another situation of interest is slab scattering, where the potential is given by a strip aligned orthogonal to the waveguide direction. We suppose the potential has the distribution
\begin{equation}
    V_{c,a}(\bm{r})=\begin{cases}
        V_0,\quad &0\leq x\leq x_0,\\
        0, &\text{elsewhere},
    \end{cases}\label{eq:slab_distribution}
\end{equation}
and the incident field takes a plane wave form
\begin{equation}
\psi_{i}=\begin{pmatrix}
    u_0\\
    v_0
\end{pmatrix}e^{\mathrm{i}(q_0x+p_0y)}, \label{eq:slab_chiral_inc}
\end{equation}
propagating from $x<0$. The field amplitudes $u_0$ and $v_0$ obey the transport equations and the wavenumbers $q_0$ and $p_0$ satisfy the eikonal equations as described in \Cref{sec:goptics}. The boundary of the slab divides the space into three regions, each of which has a uniform potential. Since the potential $V$ is independent of $y$, the total field can be expressed as
$$\psi(x,y) = \tilde{\psi}(x)e^{\mathrm{i}p_0y},$$
where $\tilde{\psi}$ is solely a function of $x$. As a result, the Dirac equation in each region reduces to an ordinary differential equation with respect to $x$, given by
\begin{equation}
    -\mathrm{i}\partial_{x}\psi-p\psi+k_{c}\bigl(\beta+V_c(x,y)\bigr)\psi=0 \label{eq:slab_chiral_dirac_ode}
\end{equation}
in the chiral array and 
\begin{equation}
\mathrm{i}\beta\partial_{x}\psi-p\psi+k_a\bigl(V_a(x,y)-1\bigr)\psi=0.
\label{eq:slab_anti_dirac_ode}
\end{equation}
in the antichiral array. The general solution to \eqref{eq:slab_chiral_dirac_ode} and \eqref{eq:slab_anti_dirac_ode} takes the form
\begin{equation}
    \psi=\begin{pmatrix}
        \phi^{+}_1\\
        \phi^{+}_2
    \end{pmatrix}e^{\mathrm{i}(q_1x+p_0y)}+\begin{pmatrix}
        \phi^{-}_1\\
        \phi^{-}_2
    \end{pmatrix}e^{\mathrm{i}(q_2x+p_0y)},\label{eq:anti_slab_general_sol}
\end{equation}
where the wavenumbers $q_{1,2}$ are given by
\begin{equation}
    q_{1,2}=-Vk_c \pm \sqrt{k_c^2+p_0^2},\quad q_{1,2}=\pm \sqrt{(V-1)^2k_a^2-p_0^2}
    \label{eq:slab_scatt_wavenumber}
\end{equation}
in the chiral and antichiral cases, respectively.  Here the potential $V$ takes the value $V_0$ inside the slab and vanishes outside. Since the transverse wavenumber $p_0$ is conserved across the interface, \eqref{eq:slab_scatt_wavenumber} implies that
\begin{equation}
    \sin\theta_i = |V_0-1|\sin \theta_t
    \label{eq: anti_snell}
\end{equation}
in the antichiral array, where $\theta_{i,t}$ denote the incident and transmitted angles from the interface normal. Eq.~\eqref{eq: anti_snell} serves as an analog of Snell's law in classical optics, which has no counterpart in the chiral array due to the absence of backscattering. The amplitudes $\phi^{\pm}_{1,2}$ are constant and coupled by the Dirac equation according to
\begin{equation}
    \phi^{\pm}_{1}=\frac{ q_{1,2}+(V_0-1)k}{p}\phi^{\pm}_{2}
    \label{eq:slab_couple}
\end{equation}
for both arrays, where $k = k_c \text{ or } k_a$. 
In the following, we examine the chiral and antichiral cases separately. In the chiral case, we recall that the $x$ coordinate is time-like, along which backscattering is prohibited. As a result, we express the field in each region as
\begin{equation}
    \psi(x,y)=\psi_i(x,y),\quad x<0,
\end{equation}
\begin{align}
    \psi(x,y)&=\begin{pmatrix}
        a\\
        b
    \end{pmatrix}e^{\mathrm{i}(q_1x+p_0y)}+\begin{pmatrix}
        c\\
        d
    \end{pmatrix}e^{\mathrm{i}(q_2x+p_0y)}, & &0<x<x_0,\\
    \psi(x,y)&=\begin{pmatrix}
        f\\
        g
    \end{pmatrix}e^{\mathrm{i}(q_0x+p_0y)}+\begin{pmatrix}
        h\\
        w
    \end{pmatrix}e^{\mathrm{i}(-q_0x+p_0y)}, & &x>x_0,\label{eq:trans}
\end{align}
where the amplitudes $\{a, b, c, d,  f, g, h, w\}$ are coupled through the Dirac equation according to \eqref{eq:slab_couple}, which yields four equations in total. The other four equations come from the matching conditions on the boundary $x=0$ and $x=x_0$, which are
\begin{equation}
\begin{cases}
    a+c=u_0,\\
    b+d=v_0,
\end{cases}
\label{eq:chiral_slab_match_0}
\end{equation}
and 
\begin{equation}
\begin{cases}
ae^{\mathrm{i}q_1x_0}+ce^{\mathrm{i}q_2x_0}=fe^{\mathrm{i}q_1x_0}+he^{\mathrm{i}q_2x_0},\\
be^{\mathrm{i}q_1x_0}+de^{\mathrm{i}q_2x_0}=ge^{\mathrm{i}q_1x_0}+we^{\mathrm{i}q_2x_0}.
\end{cases}
\label{eq:chiral_slab_match_x0}
\end{equation}
Solving Eqs.~\eqref{eq:chiral_slab_match_0} and \eqref{eq:chiral_slab_match_x0} with the aid of \eqref{eq:slab_couple}, we obtain
\begin{equation}
    a=u_0,\quad b=v_0,\quad f=u_0e^{-\mathrm{i}V_0k_cx_0},\quad g=v_0e^{-\mathrm{i}V_0k_cx_0},
\end{equation}
and $\quad c=d=h=w=0$. Comparing the transmitted field $\psi(x>x_0,y)$ in \eqref{eq:trans} with the incident field, we note that the presence of the slab in the chiral array contributes only an additional phase factor $e^{-\mathrm{i}V_0k_cx_0}$ to the field.

Next, we turn to the antichiral case, where the coordinates $x$ and $y$ are both space-like. In this setting, we apply the radiation condition at infinity. As a result, the total field in each region can be expressed as
\begin{equation}
    \begin{split}
        \psi(x<0,y)&=\psi_i+\begin{pmatrix}
            u_1\\
            v_1
        \end{pmatrix}e^{\mathrm{i}(-q_0x+p_0y)},\\
        \psi(0<x<x_0,y)&=\begin{pmatrix}
            a\\
            b
        \end{pmatrix}e^{\mathrm{i}(q_1x+p_0y)}+\begin{pmatrix}
            c\\
            d
        \end{pmatrix}e^{\mathrm{i}(q_2x+p_0y)},\\
        \psi(x>x_0,y)&=\begin{pmatrix}
            f\\
            g
        \end{pmatrix}e^{\mathrm{i}(q_0x+p_0y)},
    \end{split}
    \label{eq:anti_slab_sol_expr}
\end{equation}
where the amplitudes $(u_1, v_1)^T$ and $(f, g)^T$ correspond to the reflected and transmitted field.
The matching conditions become
\begin{equation}
\begin{cases}
    a+c=u_0+u_1,\\
    b+d=v_0+v_1,
\end{cases}
\label{eq:anti_slab_match_0}
\end{equation}
on the boundary $x=0$ and 
\begin{equation}
\begin{cases}
    ae^{\mathrm{i}q_1 x_0}+ce^{\mathrm{i}q_2 x_0}=fe^{\mathrm{i}q_0 x_0},\\
    be^{\mathrm{i}q_1 x_0}+de^{\mathrm{i}q_2 x_0}=ge^{\mathrm{i}q_0 x_0},
\end{cases}
    \label{eq:anti_slab_matching_x0}
\end{equation}
on the boundary $x=x_0$. Solving \eqref{eq:anti_slab_match_0} and \eqref{eq:anti_slab_matching_x0} with the help of \eqref{eq:slab_couple} yields
\begin{equation}
\begin{split}
    &b=\tfrac{2q_0(q_0+q_1-V_0k_a)}{(q_0+q_1)^2-(V_0k_a)^2-\bigl((q_0-q_1)^2-(V_0k_a)^2\bigr)e^{2\mathrm{i}q_1x_0}} v_0,\\
    &d=\tfrac{2q_0(q_1-q_0+V_0k_a)e^{2\mathrm{i}q_1x_0}}{(q_0+q_1)^2-(V_0k_a)^2-\bigl((q_0-q_1)^2-(V_0k_a)^2\bigr)e^{2\mathrm{i}q_1x_0}} v_0,\\
    &g=\tfrac{4q_0q_1e^{\mathrm{i}(q_1-q_0)x_0}}{(q_0+q_1)^2-(V_0k_a)^2-\bigl((q_0-q_1)^2-(V_0k_a)^2\bigr)e^{2\mathrm{i}q_1x_0}} v_0,\\
    &v_1=\left(\tfrac{2q_0\bigl(q_0+q_1-V_0k_a+(q_1-q_0+V_0k_a)e^{2\mathrm{i}q_1x_0}\bigr)}{(q_0+q_1)^2-(V_0k_a)^2-\bigl((q_0-q_1)^2-(V_0k_a)^2\bigr)e^{2\mathrm{i}q_1x_0}}-1\right)v_0.
\end{split}\label{eq:anti_slab_coeff_relation_1}
\end{equation}
The other four amplitudes $\{a,c,f,u_1\}$ can be obtained through \eqref{eq:slab_couple}:
\begin{align}
    a&=\frac{q_1+(V_0-1)k_a}{p_0}b,&
    c&=\frac{-q_1+(V_0-1)k_a}{p_0}d,\\
    f&=\frac{q_0-k_a}{p_0}g,&
    u_1&=-\frac{q_0+k_a}{p_0}v_1.
\end{align}
\begin{figure*}[t]
\hspace{-0.5cm}
     \begin{subfigure}[t]{0.25\linewidth}
     \begin{tikzpicture}
			\node[inner sep=0pt] (0,0) {
         \includegraphics[width=\textwidth]{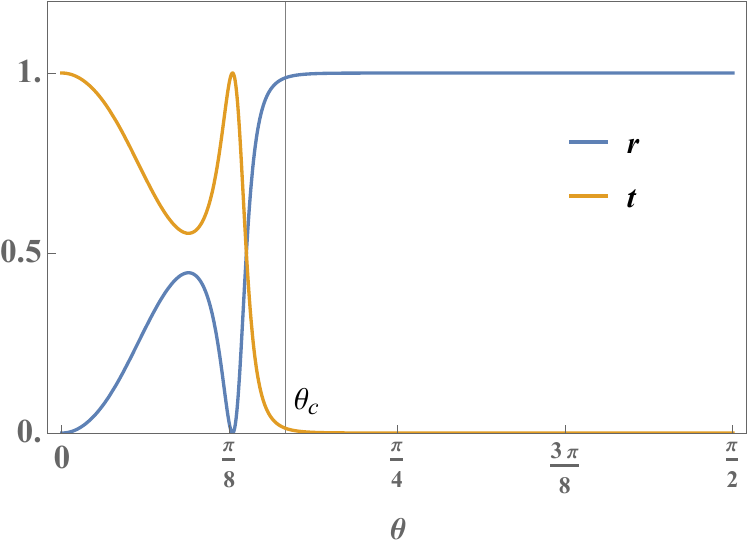}};
			\node at (-1.8,2) {(a)};
		\end{tikzpicture}
     \end{subfigure}\hfill
     \begin{subfigure}[t]{0.25\linewidth}
     \begin{tikzpicture}
			\node[inner sep=0pt] (0,0) {
         \includegraphics[width=\textwidth]{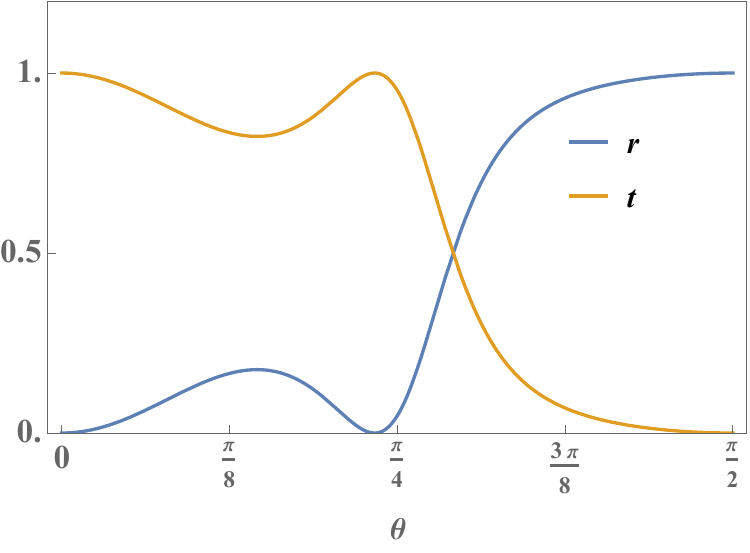}};
			\node at (-1.8,2) {(b)};
		\end{tikzpicture}
     \end{subfigure}\hfill
     \begin{subfigure}[t]{0.25\linewidth}
     \begin{tikzpicture}
			\node[inner sep=0pt] (0,0) {
         \includegraphics[width=\textwidth]{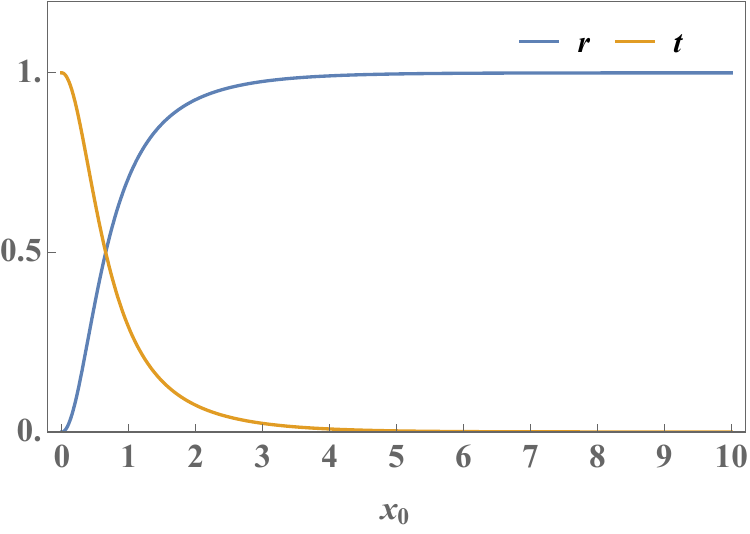}};
			\node at (-1.8,2) {(c)};
		\end{tikzpicture}
     \end{subfigure}\hfill
     \begin{subfigure}[t]{0.25\linewidth}
     \begin{tikzpicture}
			\node[inner sep=0pt] (0,0) {
         \includegraphics[width=\textwidth]{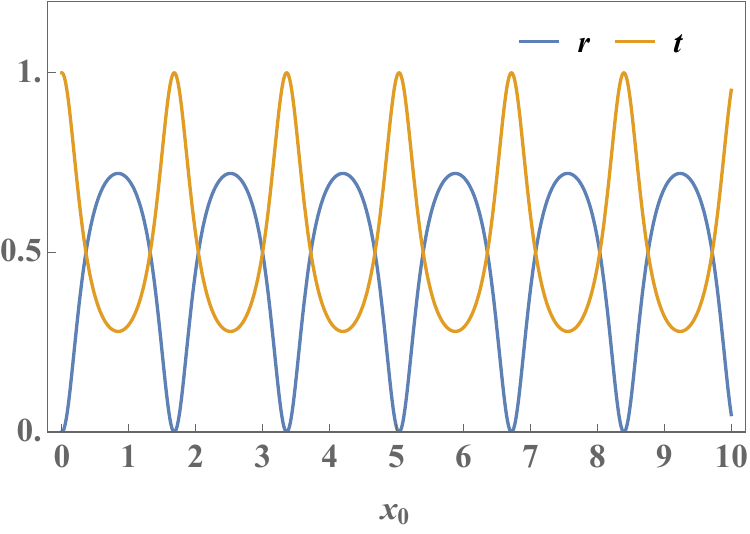}};
			\node at (-1.8,2) {(d)};
		\end{tikzpicture}
     \end{subfigure}
\caption{Reflection and transmission coefficients $r$ and $t$ of slab scattering in the antichiral array, plotted with respect to the incident angle $\theta$ in (a) with $V_0=1.5$, and (b) with $V_0=3$. In (c) and (d), we plot the scattering coefficients versus the slab width $x_0$ with $V_0=1.5$ and $V_0=3$, respectively.}
\label{fig:slab_scatt}
\end{figure*}
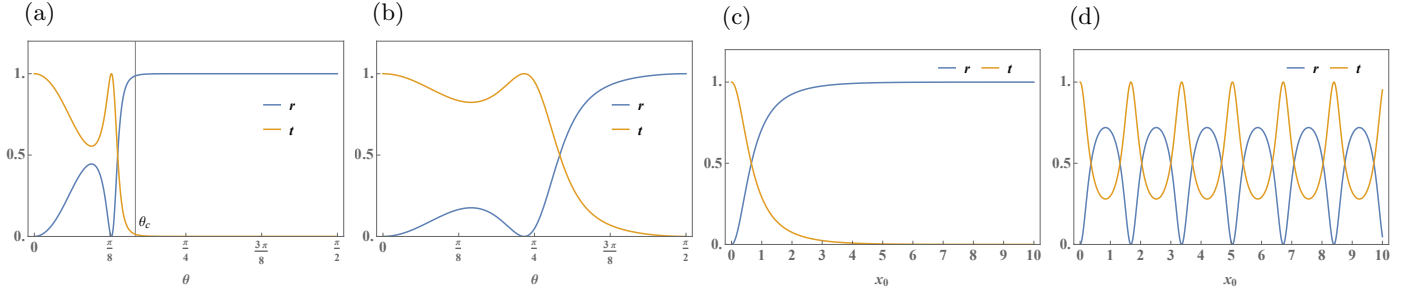
We note that the amplitudes of the reflected field $u_1, v_1$ and the transmitted field $f,g$ are both nonzero in the antichiral array. Consequently, we can define the reflection and transmission coefficients according to
\begin{equation}
    r=\frac{(\bm{j}_{\mathbf{r}})_x}{(\bm{j}_{\mathbf{inc}})_x}\quad \text{and} \quad 
    t=\frac{(\bm{j}_{\mathbf{t}})_x}{(\bm{j}_{\mathbf{inc}})_x},
\end{equation}
where the subscript $x$ denotes the component in the $x$-direction. It follows from \eqref{eq:anti_current} that the coefficients $r$ and $t$ can be computed as
\begin{equation}
\begin{split}
    r&=\frac{|u_1|^2-|v_1|^2}{|v_0|^2-|u_0|^2} =\bigg|\frac{V_0^2(q_0^2-k_a^2)}{4q_0^2q_1^2}\bigg|\bigg|\frac{e^{2\mathrm{i}q_1x_0}-1}{\frac{V_0^2k_a^2-(q_0-q_1)^2}{4q_0q_1}(e^{2\mathrm{i}q_1x_0}-1)+1}\bigg|^2,
    \\
    t&=\frac{|g|^2-|f|^2}{|v_0|^2-|u_0|^2}=\bigg|\frac{1}{\frac{V_0^2k_a^2-(q_0-q_1)^2}{4q_0q_1}(e^{2\mathrm{i}q_1x_0}-1)+1}\bigg|^2.
\end{split}
\end{equation}
We see that $r+t=1$, in agreement with conservation of probability. We plot $r$ and $t$ with respect to the incident angle $\theta$ and the slab width $x_0$ in \Cref{fig:slab_scatt} for two choices of potential $V_0$ ($1.5$ and $3$, respectively). In \Cref{fig:slab_scatt}(a), we observe total reflection when the incident angle exceeds a critical value $\theta_c$, under which the wavenumbers $q_{1,2}$ become imaginary and the field inside the slab becomes evanescent with exponential decay in the positive $x$-direction. The critical angle $\theta_c$ can be calculated as
\begin{equation}
    \theta_c=\arcsin{(V_0-1)}.
\end{equation}
If $V_0-1>1$, no critical angle exists as shown in \Cref{fig:slab_scatt}(b). In addition, we observe that the reflection and transmission coefficients become oscillatory with respect to the slab width $x_0$ as shown in \Cref{fig:slab_scatt}(d).

\section{Discussion}\label{sec: discussion}
We have considered geometrical optics, diffraction, and scattering of single photons in arrays of one-way waveguides interspersed with atoms. Throughout, we have considered two separate cases. In the chiral array, the Dirac equation is hyperbolic, with one of the coordinates time-like. As a result, scattering and diffraction feature a light-cone behavior, at which the propagated field may be discontinuous and where a scatterer has a limited domain of influence. In contrast, the antichiral Dirac equation is elliptic with both coordinates space like, leading to scattering behavior resembles classical optics. Moreover, the system possesses rotational symmetry, allowing an explicit calculation for circular scatterers. In both cases, we derive the eikonal and transport equations in geometrical optics, construct the propagator governing diffraction and compute the scattered field from scattering by point particles and slabs.

Several topics merit further study. Of particular interest is the investigation of band structure in chiral and antichiral arrays with periodically arranged atoms. In this setting, the chiral array behaves as an effective space-time periodic structure, whereas the antichiral array remains purely spatially periodic, leading to distinct spectral properties of the system. In addition, the mass terms appearing in Eqs.~\eqref{eq:chiral_dirac_nondim} and \eqref{eq:anti_dirac_nondim} may introduce topological effects in photon transport, motivating the exploration of analogs of edges states in these arrays. Finally, the present framework provides a natural extension to the two-photon regime, where the transport of entanglement is of particular interest.

\appendix
\section{Appendix: Solving the eikonal equations}\label{sec:app}
\subsection{Linear gradient} \label{sec:eik}
In this section, we present the solutions of the eikonal equations for the chiral and antichiral arrays in the presence of a linear gradient of the potential $V_{c,a}$. We first consider the chiral case, where the eikonal equation is given by \eqref{eq:chiral_eikonal}. According to \eqref{eq:general_eikonal_character}, the characteristic curves obey
\begin{equation}
    \frac{dx}{dt}=2(q+V_c),\quad \frac{dy}{dt}=-2p,\label{eq:geo_chiral_character}
\end{equation}
with the phase factor $S$ and its partial derivatives satisfying
\begin{equation}
    \frac{dS}{dt}=2q(q+V_c)-2p^2,
    \label{eq:geo_chiral_phase}
\end{equation}
and 
\begin{equation}
    \frac{dq}{dt}=-2(q+V_c)\partial_{x}V_c,\quad
    \frac{dp}{dt}=-2(q+V_c)\partial_{y}V_c.
\label{eq:geo_chiral_phase_dev}
\end{equation}
Further derivations based on \eqref{eq:geo_chiral_character} and \eqref{eq:geo_chiral_phase_dev} yield
\begin{equation}
     \frac{d^2x}{dt^2}=2\partial_{y}V_c\frac{dy}{dt},\quad
     \frac{d^2y}{dt^2}=2\partial_{y}V_c\frac{dx}{dt},
\label{eq:geo_chiral_character_ray}
\end{equation}
where the characteristic curves are related to the potential gradient.
In the case $V_c=ax$, \eqref{eq:geo_chiral_character_ray} becomes
\begin{equation}
    \begin{split}
     \frac{d^2x}{dt^2}&=2\partial_{y}V_c\frac{dy}{dt}=0,\\
     \frac{d^2y}{dt^2}&=2\partial_{y}V_c\frac{dx}{dt}=0.
\end{split}
\label{eq:chiral_ax_xy_2nd}
\end{equation}
Solving \eqref{eq:chiral_ax_xy_2nd} yields
\begin{equation}
    \begin{cases}
    x(t,s)=2q_0t+s,\\
    y(t,s)=-2p_0t,
    \end{cases}
    \label{eq:geo_chiral_ax_charact_sol}
\end{equation}
where $q_0$ and $p_0$ are constant and the characteristic curves are parameterized by the coordinates $t$ and $s$.
Substituting \eqref{eq:geo_chiral_ax_charact_sol} into \eqref{eq:geo_chiral_character}, we obtain
\begin{equation}
    q+V_c=q_0,\quad p=p_0.
    \label{eq:geo_chiral_ax_qp_expr}
\end{equation}
Using \eqref{eq:geo_chiral_ax_qp_expr}, we solve \eqref{eq:geo_chiral_phase} and obtain
\begin{equation}
\begin{split}
    S(x,y)=q_0x-\frac{a}{2}x^2+p_{0}y+S_0,
\end{split}   
\end{equation}
where $S_0$ is a constant and $q_0$ and $p_0$ are the wavenumbers satisfying
\begin{equation}
    q_0^2-p_0^2-1=0.
\end{equation}
Next, we consider the case $V_c=by$, in which \eqref{eq:geo_chiral_character_ray} becomes
\begin{equation}
    \begin{split}
     \frac{d^2x}{dt^2}&=2\partial_{y}V_c\frac{dy}{dt}=2b\frac{dy}{dt},\\
     \frac{d^2y}{dt^2}&=2\partial_{y}V_c\frac{dx}{dt}=2b\frac{dx}{dt}.
\end{split}
\label{eq:chiral_by_xy_2nd}
\end{equation}
The solution to \eqref{eq:chiral_by_xy_2nd} can be expressed as
\begin{equation}
\begin{cases}
    x(t;s)=B_1\sinh\lambda+B_2\cosh\lambda+s,\\
    y(t;s)=B_1\cosh\lambda+B_2\sinh\lambda+y_0,
\end{cases}
\label{eq:chiral_by_chracter_expr}
\end{equation}
where
\begin{equation}
    \lambda=2bt,
\end{equation}
and $B_{1,2}$ and $y_0$ are constants to be determined.
Substituting \eqref{eq:chiral_by_chracter_expr} into \eqref{eq:geo_chiral_character} yields
\begin{equation}
    q=-by_0, \quad p=-b(x-s).\label{eq:chiral_by_S_dev}
\end{equation}
Substituting \eqref{eq:chiral_by_S_dev} into the eikonal equation \eqref{eq:chiral_eikonal}, we obtain
\begin{equation}
    B_1^2-B_2^2=\frac{1}{b^2}.
    \label{eq:chiral_by_AB}
\end{equation}
Equation \eqref{eq:chiral_by_AB} suggests that the characteristic curves described by \eqref{eq:chiral_by_chracter_expr} are hyperbolic, satisfying
\begin{equation}
    (y-y_0)^2-(x-s)^2=\frac{1}{b^2}.
    \label{eq:chiral_by_character_hyperbolic}
\end{equation}
Eq.~\eqref{eq:chiral_by_character_hyperbolic} implies that the characteristic curves associated with different values of $s$ may intersect. As a result, the value $S$ at a given point $(x,y)$ could be determined by multiple characteristic curves, potentially leading to inconsistent results. To ensure uniqueness of $S(x,y)$, we thereby require $\lambda \geq \lambda_0$, in which the characteristic curves do not cross.
According to \eqref{eq:geo_chiral_phase_dev}, the phase factor $S$ along the characteristic curve can be solved as
\begin{multline}
    S(\lambda\geq \lambda_0;s)=S(\lambda_0;s)+C(\lambda_0)-by_0(B_1\sinh{\lambda}+B_2\cosh{\lambda})\\
    -b\Big(\frac{B_2^2-B_1^2}{2}\lambda+\frac{1}{2}\sinh\lambda\bigl(B_1(B_1\cosh\lambda+B_2\cosh{\lambda})\\+B_2(B_1\sinh{\lambda}+B_1\cosh{\lambda})\bigr)\Big),
\label{eq:chiral_by_phase_lambda}
\end{multline}
where $C(\lambda_0)$ is an integral constant and can be neglected.
We express the region $\lambda \geq \lambda_0$ in the Cartesian coordinates as $y\geq y_0+y_r(\lambda_0)$, where 
\begin{equation}
y_r(\lambda_0)=B_1\cosh{\lambda_0}+B_2\sinh{\lambda_0},
\end{equation}
according to \eqref{eq:chiral_by_chracter_expr}.
Consequently, \eqref{eq:chiral_by_phase_lambda} becomes
\begin{multline}
    S(x(\lambda;s),y\geq y_0+y_r)=f(s)-by_0(x-s)\\-b\Bigl(\frac{B_2^2-B_1^2}{2}\lambda(y)+\frac{1}{2}\sinh\lambda(y)\bigl(B_1(y-y_0)+B_2(x-s)\bigr)\Bigr),
    \label{eq:chiral_by_phase_xy}
\end{multline}
where $s(x,y)$ and $\lambda(y)$ are implicitly given by \eqref{eq:chiral_by_chracter_expr}. In addition, $f(s)$ denotes the phase value at the point $(x(\lambda_0;s),y_0+y_r)$ taking the form
\begin{eqnarray}
    f(s)=S(x(\lambda_0;s),y_0+y_r).
\end{eqnarray}
Substituting \eqref{eq:chiral_by_phase_xy} into \eqref{eq:chiral_by_S_dev} yields an ordinary differential equation for $f(s)$
\begin{equation}
    \frac{df(s)}{ds}=-by_0,
\end{equation}
which yields
\begin{equation}
    f(s)=-by_0s+S_0,
\end{equation}
where $S_0$ is a constant.
As a result, $S(x,y)$ in the region $y \geq y_0+y_r$ can be expressed as
\begin{multline}
S(x,y\geq y_0+y_r)=S_0-by_0x-b\Big(\frac{B_2^2-B_1^2}{2}\lambda(y)\\+\frac{1}{2}\sinh\lambda(y)\bigl(B_1(y-y_0)+B_2(x-x_0)\bigr)\Bigr).
\end{multline}
We proceed to determine $B_{1,2}$, $y_0$ and $\lambda_0$ by assuming that a plane wave is coupled to the region $y\geq y_0+y_r$ through the interface $y=y_0+y_r$. The phase factor of the plane wave can be expressed as
\begin{equation}
    S_0(x,y)=q_0x-p_0y,
\end{equation}
where
\begin{equation}
    (q_0+V_0)^2-p_0^2=1,
\end{equation}
according to the eikonal equation \eqref{eq:chiral_eikonal}. Here, the potential $V_0$ is constant in the region $y<y_0+y_r$.
If the potential $V(x,y)$ is continuous across the interface $y=y_0+y_r$, we have the matching condition at the interface
\begin{equation}
    S(x,y_0+y_r)=S_0(x,y=y_0+y_r),
    \label{eq:chiral_by_cond_phase}
\end{equation}
and 
\begin{equation}
\boldsymbol\tau_{s}\bigg|_{y=y_0+y_r}=\boldsymbol\tau_{s_0}\bigg|_{y=y_0+y_r},
\label{eq:chiral_by_cond_tan}
\end{equation}
where $\boldsymbol\tau$ is the unit tangent vector of the characteristic curve. Equations \eqref{eq:chiral_by_cond_phase} and \eqref{eq:chiral_by_cond_tan} assure the continuity of the field phase and propagation direction across the interface.
Using \eqref{eq:chiral_by_chracter_expr} and \eqref{eq:chiral_by_phase_xy}, we derive 
\begin{equation}
    y_0=-\frac{q_0}{b}, \quad \frac{B_1\sinh\lambda_0+B_2\cosh\lambda_0}{B_1\cosh\lambda_0+B_2\sinh\lambda_0}=\frac{p_0}{q_0+V_0}.
    \label{eq:chiral_by_eq_AB}
\end{equation}
In addition, the continuity of the the potential $V_c(x,y)$ across the interface gives
\begin{equation}
    V_0=b\bigl(y_0+y_r(\lambda_0)\bigr).
    \label{eq:chiral_by_index}
\end{equation}
We note that \eqref{eq:chiral_by_eq_AB} and \eqref{eq:chiral_by_index} naturally lead to \eqref{eq:chiral_by_AB}. We choose $\lambda_0=0$ for convenience and solve for $B_1$, $B_2$ based on \eqref{eq:chiral_by_eq_AB} and \eqref{eq:chiral_by_index} as
\begin{equation}
    B_1=\frac{V_0+q_0}{b},\quad B_2=\frac{p_0}{b}.
\end{equation}

Next, we consider the antichiral case where the eikonal equation is given by \eqref{eq:geo_eikonal_antichiral}. The characteristic curves are defined by
\begin{eqnarray}
     \frac{dx}{dt}=2q,\quad \frac{dy}{dt}=2p,
     \label{eq:geo_anti_character_qp}
\end{eqnarray}
where $q$ and $p$ are partial derivatives of the phase factor $S$ and we have
\begin{eqnarray}
\begin{aligned}
    &\frac{dS}{dt}=2q^2+2p^2,\\
    &
     \frac{dq}{dt}=2(V_a-1)\partial_xV_a,\quad \frac{dp}{dt}=2(V_a-1)\partial_yV_a.
     \label{eq:geo_anti_phase_qp}
\end{aligned}
\end{eqnarray}
It follows from \eqref{eq:geo_anti_character_qp} and \eqref{eq:geo_anti_phase_qp} that
\begin{eqnarray}
    \frac{d^2x}{dt^2}=4(V_a-1)\partial_xV_a,\quad
    \frac{d^2y}{dt^2}=4(V_a-1)\partial_yV_a,
    \label{eq:geo_anti_ray}
\end{eqnarray}
where the characteristic curves are related to the potential gradient. In the case $V_a=ax$, the characteristic curves solved by \eqref{eq:geo_anti_ray} take the forms
\begin{equation}
    \begin{cases}
    x(t)=A_1e^{2at}+A_2e^{-2at},\\
    y(t)=2p_1t+s,
    \end{cases}
    \label{eq:geo_anti_chracter_ax}
\end{equation}
where $A_1$, $A_2$ and $p_1$ are constant and to be determined. Upon substituting \eqref{eq:geo_anti_chracter_ax} into \eqref{eq:geo_anti_character_qp}, we obtain
\begin{eqnarray}
    \begin{cases}
q=a(A_1e^{2at}-A_2e^{-2at}),\\
p=p_1,
\end{cases}
\end{eqnarray} and
\begin{equation}
    p_1^2=4a^2A_1A_2.
\end{equation}
Similar as the chiral case, we require $t\geq t_0$ to prevent the potential intersection of characteristic curves. The phase factor $S(t;y_0)$ in the region $t\geq t_0$ is derived according to \eqref{eq:geo_anti_phase_qp} as 
\begin{equation}
    S(t\geq t_0;s)=S_0+\frac{a}{2}(A_1^2e^{4at}-A_2^2e^{-4at})+p_1^2t+p_1s,
    \label{eq:geo_anti_phase_ax}
\end{equation}
where $S_0$ is a constant. We proceed to determine $A_{1,2}$ and $p_1$ by assuming a plane wave is coupled to the region $t\geq t_0$ through the interface $t=t_0$, across which the potential is continuous. We suppose that the incident plane wave has phase $S_0(x,y)=q_0x+p_0y$ with $q_0^2+p_0^2=(V_0-1)^2$, where $V_0$ denotes the constant potential in the incident region. The matching condition at the interface can thereby be written as
\begin{equation}
\begin{split}
    S(t=t_0;s)&=S_0(x=x_0,y),\\
    \boldsymbol\tau_{s}\bigg|_{x=x_0}&=\boldsymbol\tau_{s_0}\bigg|_{x=x_0},\\
\end{split}   
\label{eq:geo_anti_ax_matching}
\end{equation}
where $\boldsymbol\tau$ is the unit tangent vector of the characteristic curve. We note that the interface $t=t_0$ is expressed as $x=x_0$ in the cartesian coordinates according to
\begin{equation}
\begin{cases}
    x_0=A_1e^{2at_0}+A_2e^{-2at_0},\\
    y_0=2p_1t_0+s.
\end{cases}
\end{equation}
We choose $t_0=0$ and solve for $p_1$, $A_1$, $A_2$ as
\begin{equation}
    p_1=p_0,\quad A_1=\frac{V_0-1+q_0}{2a},\quad A_2=\frac{V_0-1-q_0}{2a}.
    \label{eq:geo_anti_ax_coeff}
\end{equation}
For the case $V_a=by$, we can immediately obtain the result by noting that the spatial coordinates $x$ and $y$ are symmetric in \eqref{eq:geo_anti_ray}. Consequently, the characteristic curves can be defined as
\begin{eqnarray}
    \begin{cases}
    x(t)=2q_1t+s,\\
    y(t)=B_1e^{2bt}+B_2e^{-2bt},
    \end{cases}
    \label{eq:geo_anti_chracter_by}
\end{eqnarray}
where $q_1$ and $B_{1,2}$ are constants to be determined. We proceed by supposing that a plane wave with the phase $S=q_0x+p_0y$ is coupled to the region $s\geq 0$ through the interface $s=0$. Following similar lines above, we find that the phase factor $S$ takes the form
\begin{eqnarray}
    S(t;s\geq0)=S_0+\frac{b}{2}(B_1^2e^{4bt}-B_2^2e^{-4bt})+q_1^2s+q_1t,
\end{eqnarray}
with
\begin{eqnarray}
    q_1=q_0,\quad B_1=\frac{V_0+p_0}{2b},\quad B_2=\frac{V_0-p_0}{2b}.
\end{eqnarray}
\subsection{Solving the transport equations} \label{sec:transp}
In this section, we solve the transport equations for the chiral and antichiral arrays with the potential defined above. We begin with the chiral case, where the transport equations are given by \eqref{eq:chiral_geo_amplitude}. In the case $V_c=ax$, we find that 
\begin{eqnarray}
    h_c(x,y)= \frac{q+V_c+1}{p}=\frac{q_0+1}{p_0}
\end{eqnarray} 
which is constant. Therefore, the transport equations reduce to
\begin{equation}
\begin{split}
    &\frac{1}{u_0}\frac{du_0}{dt}=0,\\
    &v_0=h_cu_0,
\end{split}
\end{equation}
implying that both $u_0$ and $v_0$ are constant. Next, we consider the case $V_c=by$. Using \eqref{eq:chiral_by_phase_xy}, we find that
\begin{eqnarray}
h_c(x,y)=\frac{B_1\cosh\lambda(y)+B_2\sinh\lambda(y)+1/b}{B_1\sinh\lambda(y)+B_2\cosh\lambda(y)},
\end{eqnarray}
which is independent of $x$. As a result, we have 
\begin{eqnarray}
    \partial_xh_c(x,y)=0,\quad \partial_yh_c(x,y)=\frac{1}{B_1\sinh\lambda+B_2\cosh\lambda}.
    \label{eq:geo_chiral_transport_by}
\end{eqnarray}
Substituting \eqref{eq:geo_chiral_transport_by} into \eqref{eq:chiral_geo_amplitude}, we obtain
\begin{eqnarray}
\begin{aligned}
u_0(t\geq0;x_0)&=u_0(t=0;x_0)\Bigl[\frac{(\gamma+1)(\gamma e^{2bt}-1)}{(\gamma-1)(\gamma e^{2bt}+1)}\Bigr]^{\frac{1}{2}},\\
v_0(t\geq0;x_0)&=\frac{B_1\cosh2bt+B_2\sinh2bt+1/b}{B_1\sinh2bt+B_2\cosh2bt}\\
&\quad \times\Bigl[\frac{(\gamma+1)(\gamma e^{2bt}-1)}{(\gamma-1)(\gamma e^{2bt+1})}\Bigr]^{\frac{1}{2}}u_0(t=0;x_0),
\end{aligned}
\end{eqnarray}
where
\begin{equation}
   \gamma\equiv\sqrt{\frac{B_1+B_2}{B_1-B_2}}. 
\end{equation}
We now turn to the antichiral case, whose transport equations are given by \eqref{eq:antichiral_geo_amplitude}. In the case $V_a=ax$, it follows from \eqref{eq:geo_anti_phase_ax} that \eqref{eq:antichiral_geo_amplitude} takes the form
\begin{eqnarray}
    \begin{aligned}
    &\frac{1}{u_0}\frac{du_0}{dt}=-\frac{2aA_2}{A_1e^{4at}-A_2},\\
    &v_0=\frac{2aA_2e^{-2at}}{p_1}u_0.
\end{aligned}
\label{eq:geo_anti_transport_ax}
\end{eqnarray}
Solving \eqref{eq:geo_anti_transport_ax} yields
\begin{eqnarray}
\begin{aligned}
    &u_0(t\geq0;s)=u_0(t=0;s)\Bigl[\frac{A_1-A_2}{A_1-A_2e^{-4at}}\Bigr]^{\frac{1}{2}},\\
    &v_0(t\geq0;s)=\frac{2aA_2e^{-2at}}{p_1}\Bigl[\frac{A_1-A_2}{A_1-A_2e^{-4at}}\Bigr]^{\frac{1}{2}}u_0(t=0;s),
\end{aligned}
\end{eqnarray}
where $A_{1,2}$ and $p_1$ are given by \eqref{eq:geo_anti_ax_coeff}.
Similarly, we find that the transport equations in the case $V_a=by$ become
\begin{equation}
    \begin{split}
        &\frac{1}{u_0}\frac{du_0}{dt}=-\frac{q_0}{B_1e^{2bt}-B_2e^{-2bt}},\\
        &v_0=\frac{B_1e^{2bt}+B_2e^{-2bt}-q_0/b}{B_1e^{2bt}-B_2e^{-2bt}}u_0.
    \end{split}
    \label{eq:geo_anti_transport_by}
\end{equation}
We solve \eqref{eq:geo_anti_transport_by} and obtain
\begin{equation}
\begin{split}
u_0(t\geq0;s)&=u_0(t=0;s)\Bigl[\frac{\gamma-1}{\gamma+1}\frac{\gamma e^{2bt}+1}{\gamma e^{2bt}-1}\Bigr]^{\frac{1}{2}},\\
    v_0(t\geq0;s)&=\frac{\gamma e^{2bt}-1}{\gamma e^{2bt}+1}\Bigl[\frac{\gamma-1}{\gamma+1}\frac{\gamma e^{2bt}+1}{\gamma e^{2bt}-1}\Bigr]^{\frac{1}{2}}u_0(t=0;s),
\end{split}
\end{equation}
where
\begin{equation}
    \gamma = \sqrt{\frac{B_1}{B_2}}.
\end{equation}

\clearpage

\bibliography{waveguide}{}

\begin{thebibliography}{39}%
\makeatletter
\providecommand \@ifxundefined [1]{%
 \@ifx{#1\undefined}
}%
\providecommand \@ifnum [1]{%
 \ifnum #1\expandafter \@firstoftwo
 \else \expandafter \@secondoftwo
 \fi
}%
\providecommand \@ifx [1]{%
 \ifx #1\expandafter \@firstoftwo
 \else \expandafter \@secondoftwo
 \fi
}%
\providecommand \natexlab [1]{#1}%
\providecommand \enquote  [1]{``#1''}%
\providecommand \bibnamefont  [1]{#1}%
\providecommand \bibfnamefont [1]{#1}%
\providecommand \citenamefont [1]{#1}%
\providecommand \href@noop [0]{\@secondoftwo}%
\providecommand \href [0]{\begingroup \@sanitize@url \@href}%
\providecommand \@href[1]{\@@startlink{#1}\@@href}%
\providecommand \@@href[1]{\endgroup#1\@@endlink}%
\providecommand \@sanitize@url [0]{\catcode `\\12\catcode `\$12\catcode
  `\&12\catcode `\#12\catcode `\^12\catcode `\_12\catcode `\%12\relax}%
\providecommand \@@startlink[1]{}%
\providecommand \@@endlink[0]{}%
\providecommand \url  [0]{\begingroup\@sanitize@url \@url }%
\providecommand \@url [1]{\endgroup\@href {#1}{\urlprefix }}%
\providecommand \urlprefix  [0]{URL }%
\providecommand \Eprint [0]{\href }%
\providecommand \doibase [0]{https://doi.org/}%
\providecommand \selectlanguage [0]{\@gobble}%
\providecommand \bibinfo  [0]{\@secondoftwo}%
\providecommand \bibfield  [0]{\@secondoftwo}%
\providecommand \translation [1]{[#1]}%
\providecommand \BibitemOpen [0]{}%
\providecommand \bibitemStop [0]{}%
\providecommand \bibitemNoStop [0]{.\EOS\space}%
\providecommand \EOS [0]{\spacefactor3000\relax}%
\providecommand \BibitemShut  [1]{\csname bibitem#1\endcsname}%
\let\auto@bib@innerbib\@empty
\bibitem [{\citenamefont {Roy}\ \emph {et~al.}(2017)\citenamefont {Roy},
  \citenamefont {Wilson},\ and\ \citenamefont
  {Firstenberg}}]{roy2017colloquium}%
  \BibitemOpen
  \bibfield  {author} {\bibinfo {author} {\bibfnamefont {D.}~\bibnamefont
  {Roy}}, \bibinfo {author} {\bibfnamefont {C.~M.}\ \bibnamefont {Wilson}},\
  and\ \bibinfo {author} {\bibfnamefont {O.}~\bibnamefont {Firstenberg}},\
  }\bibfield  {title} {\bibinfo {title} {Colloquium: Strongly interacting
  photons in one-dimensional continuum},\ }\href@noop {} {\bibfield  {journal}
  {\bibinfo  {journal} {Reviews of Modern Physics}\ }\textbf {\bibinfo {volume}
  {89}},\ \bibinfo {pages} {021001} (\bibinfo {year} {2017})}\BibitemShut
  {NoStop}%
\bibitem [{\citenamefont {Sheremet}\ \emph {et~al.}(2023)\citenamefont
  {Sheremet}, \citenamefont {Petrov}, \citenamefont {Iorsh}, \citenamefont
  {Poshakinskiy},\ and\ \citenamefont {Poddubny}}]{sheremet2023waveguide}%
  \BibitemOpen
  \bibfield  {author} {\bibinfo {author} {\bibfnamefont {A.~S.}\ \bibnamefont
  {Sheremet}}, \bibinfo {author} {\bibfnamefont {M.~I.}\ \bibnamefont
  {Petrov}}, \bibinfo {author} {\bibfnamefont {I.~V.}\ \bibnamefont {Iorsh}},
  \bibinfo {author} {\bibfnamefont {A.~V.}\ \bibnamefont {Poshakinskiy}},\ and\
  \bibinfo {author} {\bibfnamefont {A.~N.}\ \bibnamefont {Poddubny}},\
  }\bibfield  {title} {\bibinfo {title} {Waveguide quantum electrodynamics:
  collective radiance and photon-photon correlations},\ }\href@noop {}
  {\bibfield  {journal} {\bibinfo  {journal} {Reviews of Modern Physics}\
  }\textbf {\bibinfo {volume} {95}},\ \bibinfo {pages} {015002} (\bibinfo
  {year} {2023})}\BibitemShut {NoStop}%
\bibitem [{\citenamefont {Gonz{\'a}lez-Tudela}\ \emph
  {et~al.}(2024)\citenamefont {Gonz{\'a}lez-Tudela}, \citenamefont {Reiserer},
  \citenamefont {Garc{\'\i}a-Ripoll},\ and\ \citenamefont
  {Garc{\'\i}a-Vidal}}]{gonzalez2024light}%
  \BibitemOpen
  \bibfield  {author} {\bibinfo {author} {\bibfnamefont {A.}~\bibnamefont
  {Gonz{\'a}lez-Tudela}}, \bibinfo {author} {\bibfnamefont {A.}~\bibnamefont
  {Reiserer}}, \bibinfo {author} {\bibfnamefont {J.~J.}\ \bibnamefont
  {Garc{\'\i}a-Ripoll}},\ and\ \bibinfo {author} {\bibfnamefont {F.~J.}\
  \bibnamefont {Garc{\'\i}a-Vidal}},\ }\bibfield  {title} {\bibinfo {title}
  {Light--matter interactions in quantum nanophotonic devices},\ }\href@noop {}
  {\bibfield  {journal} {\bibinfo  {journal} {Nature Reviews Physics}\ }\textbf
  {\bibinfo {volume} {6}},\ \bibinfo {pages} {166} (\bibinfo {year}
  {2024})}\BibitemShut {NoStop}%
\bibitem [{\citenamefont {T{\"u}rschmann}\ \emph {et~al.}(2019)\citenamefont
  {T{\"u}rschmann}, \citenamefont {Le~Jeannic}, \citenamefont {Simonsen},
  \citenamefont {Haakh}, \citenamefont {G{\"o}tzinger}, \citenamefont
  {Sandoghdar}, \citenamefont {Lodahl},\ and\ \citenamefont
  {Rotenberg}}]{turschmann2019coherent}%
  \BibitemOpen
  \bibfield  {author} {\bibinfo {author} {\bibfnamefont {P.}~\bibnamefont
  {T{\"u}rschmann}}, \bibinfo {author} {\bibfnamefont {H.}~\bibnamefont
  {Le~Jeannic}}, \bibinfo {author} {\bibfnamefont {S.~F.}\ \bibnamefont
  {Simonsen}}, \bibinfo {author} {\bibfnamefont {H.~R.}\ \bibnamefont {Haakh}},
  \bibinfo {author} {\bibfnamefont {S.}~\bibnamefont {G{\"o}tzinger}}, \bibinfo
  {author} {\bibfnamefont {V.}~\bibnamefont {Sandoghdar}}, \bibinfo {author}
  {\bibfnamefont {P.}~\bibnamefont {Lodahl}},\ and\ \bibinfo {author}
  {\bibfnamefont {N.}~\bibnamefont {Rotenberg}},\ }\bibfield  {title} {\bibinfo
  {title} {Coherent nonlinear optics of quantum emitters in nanophotonic
  waveguides},\ }\href@noop {} {\bibfield  {journal} {\bibinfo  {journal}
  {Nanophotonics}\ }\textbf {\bibinfo {volume} {8}},\ \bibinfo {pages} {1641}
  (\bibinfo {year} {2019})}\BibitemShut {NoStop}%
\bibitem [{\citenamefont {Shen}\ and\ \citenamefont
  {Fan}(2005)}]{shen2005coherent}%
  \BibitemOpen
  \bibfield  {author} {\bibinfo {author} {\bibfnamefont {J.-t.}\ \bibnamefont
  {Shen}}\ and\ \bibinfo {author} {\bibfnamefont {S.}~\bibnamefont {Fan}},\
  }\bibfield  {title} {\bibinfo {title} {Coherent photon transport from
  spontaneous emission in one-dimensional waveguides},\ }\href@noop {}
  {\bibfield  {journal} {\bibinfo  {journal} {Optics letters}\ }\textbf
  {\bibinfo {volume} {30}},\ \bibinfo {pages} {2001} (\bibinfo {year}
  {2005})}\BibitemShut {NoStop}%
\bibitem [{\citenamefont {Siampour}\ \emph {et~al.}(2023)\citenamefont
  {Siampour}, \citenamefont {O’Rourke}, \citenamefont {Brash}, \citenamefont
  {Makhonin}, \citenamefont {Dost}, \citenamefont {Hallett}, \citenamefont
  {Clarke}, \citenamefont {Patil}, \citenamefont {Skolnick},\ and\
  \citenamefont {Fox}}]{siampour2023observation}%
  \BibitemOpen
  \bibfield  {author} {\bibinfo {author} {\bibfnamefont {H.}~\bibnamefont
  {Siampour}}, \bibinfo {author} {\bibfnamefont {C.}~\bibnamefont
  {O’Rourke}}, \bibinfo {author} {\bibfnamefont {A.~J.}\ \bibnamefont
  {Brash}}, \bibinfo {author} {\bibfnamefont {M.~N.}\ \bibnamefont {Makhonin}},
  \bibinfo {author} {\bibfnamefont {R.}~\bibnamefont {Dost}}, \bibinfo {author}
  {\bibfnamefont {D.~J.}\ \bibnamefont {Hallett}}, \bibinfo {author}
  {\bibfnamefont {E.}~\bibnamefont {Clarke}}, \bibinfo {author} {\bibfnamefont
  {P.~K.}\ \bibnamefont {Patil}}, \bibinfo {author} {\bibfnamefont {M.~S.}\
  \bibnamefont {Skolnick}},\ and\ \bibinfo {author} {\bibfnamefont {A.~M.}\
  \bibnamefont {Fox}},\ }\bibfield  {title} {\bibinfo {title} {Observation of
  large spontaneous emission rate enhancement of quantum dots in a
  broken-symmetry slow-light waveguide},\ }\href@noop {} {\bibfield  {journal}
  {\bibinfo  {journal} {npj Quantum Information}\ }\textbf {\bibinfo {volume}
  {9}},\ \bibinfo {pages} {15} (\bibinfo {year} {2023})}\BibitemShut {NoStop}%
\bibitem [{\citenamefont {Mitsch}\ \emph {et~al.}(2014)\citenamefont {Mitsch},
  \citenamefont {Sayrin}, \citenamefont {Albrecht}, \citenamefont
  {Schneeweiss},\ and\ \citenamefont {Rauschenbeutel}}]{mitsch2014quantum}%
  \BibitemOpen
  \bibfield  {author} {\bibinfo {author} {\bibfnamefont {R.}~\bibnamefont
  {Mitsch}}, \bibinfo {author} {\bibfnamefont {C.}~\bibnamefont {Sayrin}},
  \bibinfo {author} {\bibfnamefont {B.}~\bibnamefont {Albrecht}}, \bibinfo
  {author} {\bibfnamefont {P.}~\bibnamefont {Schneeweiss}},\ and\ \bibinfo
  {author} {\bibfnamefont {A.}~\bibnamefont {Rauschenbeutel}},\ }\bibfield
  {title} {\bibinfo {title} {Quantum state-controlled directional spontaneous
  emission of photons into a nanophotonic waveguide},\ }\href@noop {}
  {\bibfield  {journal} {\bibinfo  {journal} {Nature communications}\ }\textbf
  {\bibinfo {volume} {5}},\ \bibinfo {pages} {5713} (\bibinfo {year}
  {2014})}\BibitemShut {NoStop}%
\bibitem [{\citenamefont {Shen}\ and\ \citenamefont
  {Fan}(2007)}]{PhysRevLett.98.153003}%
  \BibitemOpen
  \bibfield  {author} {\bibinfo {author} {\bibfnamefont {J.-T.}\ \bibnamefont
  {Shen}}\ and\ \bibinfo {author} {\bibfnamefont {S.}~\bibnamefont {Fan}},\
  }\bibfield  {title} {\bibinfo {title} {Strongly correlated two-photon
  transport in a one-dimensional waveguide coupled to a two-level system},\
  }\href {https://doi.org/10.1103/PhysRevLett.98.153003} {\bibfield  {journal}
  {\bibinfo  {journal} {Phys. Rev. Lett.}\ }\textbf {\bibinfo {volume} {98}},\
  \bibinfo {pages} {153003} (\bibinfo {year} {2007})}\BibitemShut {NoStop}%
\bibitem [{\citenamefont {Mahmoodian}\ \emph {et~al.}(2018)\citenamefont
  {Mahmoodian}, \citenamefont {\ifmmode~\check{C}\else \v{C}\fi{}epulkovskis},
  \citenamefont {Das}, \citenamefont {Lodahl}, \citenamefont {Hammerer},\ and\
  \citenamefont {S\o{}rensen}}]{PhysRevLett.121.143601}%
  \BibitemOpen
  \bibfield  {author} {\bibinfo {author} {\bibfnamefont {S.}~\bibnamefont
  {Mahmoodian}}, \bibinfo {author} {\bibfnamefont {M.}~\bibnamefont
  {\ifmmode~\check{C}\else \v{C}\fi{}epulkovskis}}, \bibinfo {author}
  {\bibfnamefont {S.}~\bibnamefont {Das}}, \bibinfo {author} {\bibfnamefont
  {P.}~\bibnamefont {Lodahl}}, \bibinfo {author} {\bibfnamefont
  {K.}~\bibnamefont {Hammerer}},\ and\ \bibinfo {author} {\bibfnamefont
  {A.~S.}\ \bibnamefont {S\o{}rensen}},\ }\bibfield  {title} {\bibinfo {title}
  {Strongly correlated photon transport in waveguide quantum electrodynamics
  with weakly coupled emitters},\ }\href
  {https://doi.org/10.1103/PhysRevLett.121.143601} {\bibfield  {journal}
  {\bibinfo  {journal} {Phys. Rev. Lett.}\ }\textbf {\bibinfo {volume} {121}},\
  \bibinfo {pages} {143601} (\bibinfo {year} {2018})}\BibitemShut {NoStop}%
\bibitem [{\citenamefont {Masson}\ and\ \citenamefont
  {Asenjo-Garcia}(2020)}]{PhysRevResearch.2.043213}%
  \BibitemOpen
  \bibfield  {author} {\bibinfo {author} {\bibfnamefont {S.~J.}\ \bibnamefont
  {Masson}}\ and\ \bibinfo {author} {\bibfnamefont {A.}~\bibnamefont
  {Asenjo-Garcia}},\ }\bibfield  {title} {\bibinfo {title} {Atomic-waveguide
  quantum electrodynamics},\ }\href
  {https://doi.org/10.1103/PhysRevResearch.2.043213} {\bibfield  {journal}
  {\bibinfo  {journal} {Phys. Rev. Res.}\ }\textbf {\bibinfo {volume} {2}},\
  \bibinfo {pages} {043213} (\bibinfo {year} {2020})}\BibitemShut {NoStop}%
\bibitem [{\citenamefont {Solano}\ \emph {et~al.}(2017)\citenamefont {Solano},
  \citenamefont {Barberis-Blostein}, \citenamefont {Fatemi}, \citenamefont
  {Orozco},\ and\ \citenamefont {Rolston}}]{solano2017super}%
  \BibitemOpen
  \bibfield  {author} {\bibinfo {author} {\bibfnamefont {P.}~\bibnamefont
  {Solano}}, \bibinfo {author} {\bibfnamefont {P.}~\bibnamefont
  {Barberis-Blostein}}, \bibinfo {author} {\bibfnamefont {F.~K.}\ \bibnamefont
  {Fatemi}}, \bibinfo {author} {\bibfnamefont {L.~A.}\ \bibnamefont {Orozco}},\
  and\ \bibinfo {author} {\bibfnamefont {S.~L.}\ \bibnamefont {Rolston}},\
  }\bibfield  {title} {\bibinfo {title} {Super-radiance reveals infinite-range
  dipole interactions through a nanofiber},\ }\href@noop {} {\bibfield
  {journal} {\bibinfo  {journal} {Nature communications}\ }\textbf {\bibinfo
  {volume} {8}},\ \bibinfo {pages} {1857} (\bibinfo {year} {2017})}\BibitemShut
  {NoStop}%
\bibitem [{\citenamefont {Kim}\ \emph {et~al.}(2018)\citenamefont {Kim},
  \citenamefont {Aghaeimeibodi}, \citenamefont {Richardson}, \citenamefont
  {Leavitt},\ and\ \citenamefont {Waks}}]{kim2018super}%
  \BibitemOpen
  \bibfield  {author} {\bibinfo {author} {\bibfnamefont {J.-H.}\ \bibnamefont
  {Kim}}, \bibinfo {author} {\bibfnamefont {S.}~\bibnamefont {Aghaeimeibodi}},
  \bibinfo {author} {\bibfnamefont {C.~J.}\ \bibnamefont {Richardson}},
  \bibinfo {author} {\bibfnamefont {R.~P.}\ \bibnamefont {Leavitt}},\ and\
  \bibinfo {author} {\bibfnamefont {E.}~\bibnamefont {Waks}},\ }\bibfield
  {title} {\bibinfo {title} {Super-radiant emission from quantum dots in a
  nanophotonic waveguide},\ }\href@noop {} {\bibfield  {journal} {\bibinfo
  {journal} {Nano Letters}\ }\textbf {\bibinfo {volume} {18}},\ \bibinfo
  {pages} {4734} (\bibinfo {year} {2018})}\BibitemShut {NoStop}%
\bibitem [{\citenamefont {Akhlaghi}\ \emph {et~al.}(2015)\citenamefont
  {Akhlaghi}, \citenamefont {Schelew},\ and\ \citenamefont
  {Young}}]{akhlaghi2015waveguide}%
  \BibitemOpen
  \bibfield  {author} {\bibinfo {author} {\bibfnamefont {M.~K.}\ \bibnamefont
  {Akhlaghi}}, \bibinfo {author} {\bibfnamefont {E.}~\bibnamefont {Schelew}},\
  and\ \bibinfo {author} {\bibfnamefont {J.~F.}\ \bibnamefont {Young}},\
  }\bibfield  {title} {\bibinfo {title} {Waveguide integrated superconducting
  single-photon detectors implemented as near-perfect absorbers of coherent
  radiation},\ }\href@noop {} {\bibfield  {journal} {\bibinfo  {journal}
  {Nature communications}\ }\textbf {\bibinfo {volume} {6}},\ \bibinfo {pages}
  {8233} (\bibinfo {year} {2015})}\BibitemShut {NoStop}%
\bibitem [{\citenamefont {Politi}\ \emph {et~al.}(2008)\citenamefont {Politi},
  \citenamefont {Cryan}, \citenamefont {Rarity}, \citenamefont {Yu},\ and\
  \citenamefont {O'brien}}]{politi2008silica}%
  \BibitemOpen
  \bibfield  {author} {\bibinfo {author} {\bibfnamefont {A.}~\bibnamefont
  {Politi}}, \bibinfo {author} {\bibfnamefont {M.~J.}\ \bibnamefont {Cryan}},
  \bibinfo {author} {\bibfnamefont {J.~G.}\ \bibnamefont {Rarity}}, \bibinfo
  {author} {\bibfnamefont {S.}~\bibnamefont {Yu}},\ and\ \bibinfo {author}
  {\bibfnamefont {J.~L.}\ \bibnamefont {O'brien}},\ }\bibfield  {title}
  {\bibinfo {title} {Silica-on-silicon waveguide quantum circuits},\
  }\href@noop {} {\bibfield  {journal} {\bibinfo  {journal} {Science}\ }\textbf
  {\bibinfo {volume} {320}},\ \bibinfo {pages} {646} (\bibinfo {year}
  {2008})}\BibitemShut {NoStop}%
\bibitem [{\citenamefont {Wang}\ \emph {et~al.}(2020)\citenamefont {Wang},
  \citenamefont {Sciarrino}, \citenamefont {Laing},\ and\ \citenamefont
  {Thompson}}]{wang2020integrated}%
  \BibitemOpen
  \bibfield  {author} {\bibinfo {author} {\bibfnamefont {J.}~\bibnamefont
  {Wang}}, \bibinfo {author} {\bibfnamefont {F.}~\bibnamefont {Sciarrino}},
  \bibinfo {author} {\bibfnamefont {A.}~\bibnamefont {Laing}},\ and\ \bibinfo
  {author} {\bibfnamefont {M.~G.}\ \bibnamefont {Thompson}},\ }\bibfield
  {title} {\bibinfo {title} {Integrated photonic quantum technologies},\
  }\href@noop {} {\bibfield  {journal} {\bibinfo  {journal} {Nature photonics}\
  }\textbf {\bibinfo {volume} {14}},\ \bibinfo {pages} {273} (\bibinfo {year}
  {2020})}\BibitemShut {NoStop}%
\bibitem [{\citenamefont {Lodahl}\ \emph {et~al.}(2017)\citenamefont {Lodahl},
  \citenamefont {Mahmoodian}, \citenamefont {Stobbe}, \citenamefont
  {Rauschenbeutel}, \citenamefont {Schneeweiss}, \citenamefont {Volz},
  \citenamefont {Pichler},\ and\ \citenamefont {Zoller}}]{lodahl2017chiral}%
  \BibitemOpen
  \bibfield  {author} {\bibinfo {author} {\bibfnamefont {P.}~\bibnamefont
  {Lodahl}}, \bibinfo {author} {\bibfnamefont {S.}~\bibnamefont {Mahmoodian}},
  \bibinfo {author} {\bibfnamefont {S.}~\bibnamefont {Stobbe}}, \bibinfo
  {author} {\bibfnamefont {A.}~\bibnamefont {Rauschenbeutel}}, \bibinfo
  {author} {\bibfnamefont {P.}~\bibnamefont {Schneeweiss}}, \bibinfo {author}
  {\bibfnamefont {J.}~\bibnamefont {Volz}}, \bibinfo {author} {\bibfnamefont
  {H.}~\bibnamefont {Pichler}},\ and\ \bibinfo {author} {\bibfnamefont
  {P.}~\bibnamefont {Zoller}},\ }\bibfield  {title} {\bibinfo {title} {Chiral
  quantum optics},\ }\href@noop {} {\bibfield  {journal} {\bibinfo  {journal}
  {Nature}\ }\textbf {\bibinfo {volume} {541}},\ \bibinfo {pages} {473}
  (\bibinfo {year} {2017})}\BibitemShut {NoStop}%
\bibitem [{\citenamefont {Su\'arez-Forero}\ \emph {et~al.}(2025)\citenamefont
  {Su\'arez-Forero}, \citenamefont {Jalali~Mehrabad}, \citenamefont {Vega},
  \citenamefont {Gonz\'alez-Tudela},\ and\ \citenamefont
  {Hafezi}}]{PRXQuantum.6.020101}%
  \BibitemOpen
  \bibfield  {author} {\bibinfo {author} {\bibfnamefont {D.}~\bibnamefont
  {Su\'arez-Forero}}, \bibinfo {author} {\bibfnamefont {M.}~\bibnamefont
  {Jalali~Mehrabad}}, \bibinfo {author} {\bibfnamefont {C.}~\bibnamefont
  {Vega}}, \bibinfo {author} {\bibfnamefont {A.}~\bibnamefont
  {Gonz\'alez-Tudela}},\ and\ \bibinfo {author} {\bibfnamefont
  {M.}~\bibnamefont {Hafezi}},\ }\bibfield  {title} {\bibinfo {title} {Chiral
  quantum optics: Recent developments and future directions},\ }\href
  {https://doi.org/10.1103/PRXQuantum.6.020101} {\bibfield  {journal} {\bibinfo
   {journal} {PRX Quantum}\ }\textbf {\bibinfo {volume} {6}},\ \bibinfo {pages}
  {020101} (\bibinfo {year} {2025})}\BibitemShut {NoStop}%
\bibitem [{\citenamefont {Mahmoodian}\ \emph {et~al.}(2016)\citenamefont
  {Mahmoodian}, \citenamefont {Lodahl},\ and\ \citenamefont
  {S\o{}rensen}}]{PhysRevLett.117.240501}%
  \BibitemOpen
  \bibfield  {author} {\bibinfo {author} {\bibfnamefont {S.}~\bibnamefont
  {Mahmoodian}}, \bibinfo {author} {\bibfnamefont {P.}~\bibnamefont {Lodahl}},\
  and\ \bibinfo {author} {\bibfnamefont {A.~S.}\ \bibnamefont {S\o{}rensen}},\
  }\bibfield  {title} {\bibinfo {title} {Quantum networks with
  chiral-light--matter interaction in waveguides},\ }\href
  {https://doi.org/10.1103/PhysRevLett.117.240501} {\bibfield  {journal}
  {\bibinfo  {journal} {Phys. Rev. Lett.}\ }\textbf {\bibinfo {volume} {117}},\
  \bibinfo {pages} {240501} (\bibinfo {year} {2016})}\BibitemShut {NoStop}%
\bibitem [{\citenamefont {Gonzalez-Ballestero}\ \emph
  {et~al.}(2016)\citenamefont {Gonzalez-Ballestero}, \citenamefont {Moreno},
  \citenamefont {Garcia-Vidal},\ and\ \citenamefont
  {Gonzalez-Tudela}}]{PhysRevA.94.063817}%
  \BibitemOpen
  \bibfield  {author} {\bibinfo {author} {\bibfnamefont {C.}~\bibnamefont
  {Gonzalez-Ballestero}}, \bibinfo {author} {\bibfnamefont {E.}~\bibnamefont
  {Moreno}}, \bibinfo {author} {\bibfnamefont {F.~J.}\ \bibnamefont
  {Garcia-Vidal}},\ and\ \bibinfo {author} {\bibfnamefont {A.}~\bibnamefont
  {Gonzalez-Tudela}},\ }\bibfield  {title} {\bibinfo {title} {Nonreciprocal
  few-photon routing schemes based on chiral waveguide-emitter couplings},\
  }\href {https://doi.org/10.1103/PhysRevA.94.063817} {\bibfield  {journal}
  {\bibinfo  {journal} {Phys. Rev. A}\ }\textbf {\bibinfo {volume} {94}},\
  \bibinfo {pages} {063817} (\bibinfo {year} {2016})}\BibitemShut {NoStop}%
\bibitem [{\citenamefont {Poudyal}\ and\ \citenamefont
  {Mirza}(2020)}]{PhysRevResearch.2.043048}%
  \BibitemOpen
  \bibfield  {author} {\bibinfo {author} {\bibfnamefont {B.}~\bibnamefont
  {Poudyal}}\ and\ \bibinfo {author} {\bibfnamefont {I.~M.}\ \bibnamefont
  {Mirza}},\ }\bibfield  {title} {\bibinfo {title} {Collective photon routing
  improvement in a dissipative quantum emitter chain strongly coupled to a
  chiral waveguide qed ladder},\ }\href
  {https://doi.org/10.1103/PhysRevResearch.2.043048} {\bibfield  {journal}
  {\bibinfo  {journal} {Phys. Rev. Res.}\ }\textbf {\bibinfo {volume} {2}},\
  \bibinfo {pages} {043048} (\bibinfo {year} {2020})}\BibitemShut {NoStop}%
\bibitem [{\citenamefont {Petersen}\ \emph {et~al.}(2014)\citenamefont
  {Petersen}, \citenamefont {Volz},\ and\ \citenamefont
  {Rauschenbeutel}}]{petersen2014chiral}%
  \BibitemOpen
  \bibfield  {author} {\bibinfo {author} {\bibfnamefont {J.}~\bibnamefont
  {Petersen}}, \bibinfo {author} {\bibfnamefont {J.}~\bibnamefont {Volz}},\
  and\ \bibinfo {author} {\bibfnamefont {A.}~\bibnamefont {Rauschenbeutel}},\
  }\bibfield  {title} {\bibinfo {title} {Chiral nanophotonic waveguide
  interface based on spin-orbit interaction of light},\ }\href@noop {}
  {\bibfield  {journal} {\bibinfo  {journal} {Science}\ }\textbf {\bibinfo
  {volume} {346}},\ \bibinfo {pages} {67} (\bibinfo {year} {2014})}\BibitemShut
  {NoStop}%
\bibitem [{\citenamefont {S{\"o}llner}\ \emph {et~al.}(2015)\citenamefont
  {S{\"o}llner}, \citenamefont {Mahmoodian}, \citenamefont {Hansen},
  \citenamefont {Midolo}, \citenamefont {Javadi}, \citenamefont
  {Kir{\v{s}}ansk{\.e}}, \citenamefont {Pregnolato}, \citenamefont {El-Ella},
  \citenamefont {Lee}, \citenamefont {Song} \emph
  {et~al.}}]{sollner2015deterministic}%
  \BibitemOpen
  \bibfield  {author} {\bibinfo {author} {\bibfnamefont {I.}~\bibnamefont
  {S{\"o}llner}}, \bibinfo {author} {\bibfnamefont {S.}~\bibnamefont
  {Mahmoodian}}, \bibinfo {author} {\bibfnamefont {S.~L.}\ \bibnamefont
  {Hansen}}, \bibinfo {author} {\bibfnamefont {L.}~\bibnamefont {Midolo}},
  \bibinfo {author} {\bibfnamefont {A.}~\bibnamefont {Javadi}}, \bibinfo
  {author} {\bibfnamefont {G.}~\bibnamefont {Kir{\v{s}}ansk{\.e}}}, \bibinfo
  {author} {\bibfnamefont {T.}~\bibnamefont {Pregnolato}}, \bibinfo {author}
  {\bibfnamefont {H.}~\bibnamefont {El-Ella}}, \bibinfo {author} {\bibfnamefont
  {E.~H.}\ \bibnamefont {Lee}}, \bibinfo {author} {\bibfnamefont {J.~D.}\
  \bibnamefont {Song}}, \emph {et~al.},\ }\bibfield  {title} {\bibinfo {title}
  {Deterministic photon--emitter coupling in chiral photonic circuits},\
  }\href@noop {} {\bibfield  {journal} {\bibinfo  {journal} {Nature
  nanotechnology}\ }\textbf {\bibinfo {volume} {10}},\ \bibinfo {pages} {775}
  (\bibinfo {year} {2015})}\BibitemShut {NoStop}%
\bibitem [{\citenamefont {Coles}\ \emph {et~al.}(2016)\citenamefont {Coles},
  \citenamefont {Price}, \citenamefont {Dixon}, \citenamefont {Royall},
  \citenamefont {Clarke}, \citenamefont {Kok}, \citenamefont {Skolnick},
  \citenamefont {Fox},\ and\ \citenamefont {Makhonin}}]{coles2016chirality}%
  \BibitemOpen
  \bibfield  {author} {\bibinfo {author} {\bibfnamefont {R.}~\bibnamefont
  {Coles}}, \bibinfo {author} {\bibfnamefont {D.}~\bibnamefont {Price}},
  \bibinfo {author} {\bibfnamefont {J.}~\bibnamefont {Dixon}}, \bibinfo
  {author} {\bibfnamefont {B.}~\bibnamefont {Royall}}, \bibinfo {author}
  {\bibfnamefont {E.}~\bibnamefont {Clarke}}, \bibinfo {author} {\bibfnamefont
  {P.}~\bibnamefont {Kok}}, \bibinfo {author} {\bibfnamefont {M.}~\bibnamefont
  {Skolnick}}, \bibinfo {author} {\bibfnamefont {A.}~\bibnamefont {Fox}},\ and\
  \bibinfo {author} {\bibfnamefont {M.}~\bibnamefont {Makhonin}},\ }\bibfield
  {title} {\bibinfo {title} {Chirality of nanophotonic waveguide with embedded
  quantum emitter for unidirectional spin transfer},\ }\href@noop {} {\bibfield
   {journal} {\bibinfo  {journal} {Nature communications}\ }\textbf {\bibinfo
  {volume} {7}},\ \bibinfo {pages} {11183} (\bibinfo {year}
  {2016})}\BibitemShut {NoStop}%
\bibitem [{\citenamefont {Mirza}\ \emph {et~al.}(2017)\citenamefont {Mirza},
  \citenamefont {Hoskins},\ and\ \citenamefont
  {Schotland}}]{mirza2017chirality}%
  \BibitemOpen
  \bibfield  {author} {\bibinfo {author} {\bibfnamefont {I.~M.}\ \bibnamefont
  {Mirza}}, \bibinfo {author} {\bibfnamefont {J.~G.}\ \bibnamefont {Hoskins}},\
  and\ \bibinfo {author} {\bibfnamefont {J.~C.}\ \bibnamefont {Schotland}},\
  }\bibfield  {title} {\bibinfo {title} {Chirality, band structure, and
  localization in waveguide quantum electrodynamics},\ }\href@noop {}
  {\bibfield  {journal} {\bibinfo  {journal} {Physical Review A}\ }\textbf
  {\bibinfo {volume} {96}},\ \bibinfo {pages} {053804} (\bibinfo {year}
  {2017})}\BibitemShut {NoStop}%
\bibitem [{\citenamefont {Mirza}\ and\ \citenamefont
  {Schotland}(2018)}]{mirza2018influence}%
  \BibitemOpen
  \bibfield  {author} {\bibinfo {author} {\bibfnamefont {I.~M.}\ \bibnamefont
  {Mirza}}\ and\ \bibinfo {author} {\bibfnamefont {J.~C.}\ \bibnamefont
  {Schotland}},\ }\bibfield  {title} {\bibinfo {title} {Influence of disorder
  on electromagnetically induced transparency in chiral waveguide quantum
  electrodynamics},\ }\href@noop {} {\bibfield  {journal} {\bibinfo  {journal}
  {JOSA B}\ }\textbf {\bibinfo {volume} {35}},\ \bibinfo {pages} {1149}
  (\bibinfo {year} {2018})}\BibitemShut {NoStop}%
\bibitem [{\citenamefont {Hoskins}\ \emph {et~al.}(2023)\citenamefont
  {Hoskins}, \citenamefont {Rachh},\ and\ \citenamefont
  {Schotland}}]{hoskins2023quantum}%
  \BibitemOpen
  \bibfield  {author} {\bibinfo {author} {\bibfnamefont {J.~G.}\ \bibnamefont
  {Hoskins}}, \bibinfo {author} {\bibfnamefont {M.}~\bibnamefont {Rachh}},\
  and\ \bibinfo {author} {\bibfnamefont {J.~C.}\ \bibnamefont {Schotland}},\
  }\bibfield  {title} {\bibinfo {title} {Quantum electrodynamics of chiral and
  antichiral waveguide arrays},\ }\href@noop {} {\bibfield  {journal} {\bibinfo
   {journal} {Optics Letters}\ }\textbf {\bibinfo {volume} {48}},\ \bibinfo
  {pages} {1232} (\bibinfo {year} {2023})}\BibitemShut {NoStop}%
\bibitem [{\citenamefont {Yin}\ \emph {et~al.}(2022)\citenamefont {Yin},
  \citenamefont {Galiffi},\ and\ \citenamefont {Al{\`u}}}]{yin2022floquet}%
  \BibitemOpen
  \bibfield  {author} {\bibinfo {author} {\bibfnamefont {S.}~\bibnamefont
  {Yin}}, \bibinfo {author} {\bibfnamefont {E.}~\bibnamefont {Galiffi}},\ and\
  \bibinfo {author} {\bibfnamefont {A.}~\bibnamefont {Al{\`u}}},\ }\bibfield
  {title} {\bibinfo {title} {Floquet metamaterials},\ }\href@noop {} {\bibfield
   {journal} {\bibinfo  {journal} {ELight}\ }\textbf {\bibinfo {volume} {2}},\
  \bibinfo {pages} {1} (\bibinfo {year} {2022})}\BibitemShut {NoStop}%
\bibitem [{\citenamefont {Cullen}(1958)}]{cullen1958travelling}%
  \BibitemOpen
  \bibfield  {author} {\bibinfo {author} {\bibfnamefont {A.}~\bibnamefont
  {Cullen}},\ }\bibfield  {title} {\bibinfo {title} {A travelling-wave
  parametric amplifier},\ }\href@noop {} {\bibfield  {journal} {\bibinfo
  {journal} {Nature}\ }\textbf {\bibinfo {volume} {181}},\ \bibinfo {pages}
  {332} (\bibinfo {year} {1958})}\BibitemShut {NoStop}%
\bibitem [{\citenamefont {Raiford}(1974)}]{raiford1974degenerate}%
  \BibitemOpen
  \bibfield  {author} {\bibinfo {author} {\bibfnamefont {M.}~\bibnamefont
  {Raiford}},\ }\bibfield  {title} {\bibinfo {title} {Degenerate parametric
  amplification with time-dependent pump amplitude and phase},\ }\href@noop {}
  {\bibfield  {journal} {\bibinfo  {journal} {Physical Review A}\ }\textbf
  {\bibinfo {volume} {9}},\ \bibinfo {pages} {2060} (\bibinfo {year}
  {1974})}\BibitemShut {NoStop}%
\bibitem [{\citenamefont {Fleury}\ \emph {et~al.}(2016)\citenamefont {Fleury},
  \citenamefont {Khanikaev},\ and\ \citenamefont
  {Al{\`u}}}]{fleury2016floquet}%
  \BibitemOpen
  \bibfield  {author} {\bibinfo {author} {\bibfnamefont {R.}~\bibnamefont
  {Fleury}}, \bibinfo {author} {\bibfnamefont {A.~B.}\ \bibnamefont
  {Khanikaev}},\ and\ \bibinfo {author} {\bibfnamefont {A.}~\bibnamefont
  {Al{\`u}}},\ }\bibfield  {title} {\bibinfo {title} {Floquet topological
  insulators for sound},\ }\href@noop {} {\bibfield  {journal} {\bibinfo
  {journal} {Nature communications}\ }\textbf {\bibinfo {volume} {7}},\
  \bibinfo {pages} {1} (\bibinfo {year} {2016})}\BibitemShut {NoStop}%
\bibitem [{\citenamefont {Rechtsman}\ \emph {et~al.}(2013)\citenamefont
  {Rechtsman}, \citenamefont {Zeuner}, \citenamefont {Plotnik}, \citenamefont
  {Lumer}, \citenamefont {Podolsky}, \citenamefont {Dreisow}, \citenamefont
  {Nolte}, \citenamefont {Segev},\ and\ \citenamefont
  {Szameit}}]{rechtsman2013photonic}%
  \BibitemOpen
  \bibfield  {author} {\bibinfo {author} {\bibfnamefont {M.~C.}\ \bibnamefont
  {Rechtsman}}, \bibinfo {author} {\bibfnamefont {J.~M.}\ \bibnamefont
  {Zeuner}}, \bibinfo {author} {\bibfnamefont {Y.}~\bibnamefont {Plotnik}},
  \bibinfo {author} {\bibfnamefont {Y.}~\bibnamefont {Lumer}}, \bibinfo
  {author} {\bibfnamefont {D.}~\bibnamefont {Podolsky}}, \bibinfo {author}
  {\bibfnamefont {F.}~\bibnamefont {Dreisow}}, \bibinfo {author} {\bibfnamefont
  {S.}~\bibnamefont {Nolte}}, \bibinfo {author} {\bibfnamefont
  {M.}~\bibnamefont {Segev}},\ and\ \bibinfo {author} {\bibfnamefont
  {A.}~\bibnamefont {Szameit}},\ }\bibfield  {title} {\bibinfo {title}
  {Photonic floquet topological insulators},\ }\href@noop {} {\bibfield
  {journal} {\bibinfo  {journal} {Nature}\ }\textbf {\bibinfo {volume} {496}},\
  \bibinfo {pages} {196} (\bibinfo {year} {2013})}\BibitemShut {NoStop}%
\bibitem [{\citenamefont {Parzen}(1950)}]{parzen1950scattering}%
  \BibitemOpen
  \bibfield  {author} {\bibinfo {author} {\bibfnamefont {G.}~\bibnamefont
  {Parzen}},\ }\bibfield  {title} {\bibinfo {title} {On the scattering theory
  of the dirac equation},\ }\href@noop {} {\bibfield  {journal} {\bibinfo
  {journal} {Physical Review}\ }\textbf {\bibinfo {volume} {80}},\ \bibinfo
  {pages} {261} (\bibinfo {year} {1950})}\BibitemShut {NoStop}%
\bibitem [{\citenamefont {Scheck}\ and\ \citenamefont
  {Stingl}(1968)}]{scheck1968approximate}%
  \BibitemOpen
  \bibfield  {author} {\bibinfo {author} {\bibfnamefont {F.}~\bibnamefont
  {Scheck}}\ and\ \bibinfo {author} {\bibfnamefont {M.}~\bibnamefont
  {Stingl}},\ }\bibfield  {title} {\bibinfo {title} {Approximate scattering
  solutions of the dirac equation for electron-nucleus processes in light
  nuclei},\ }\href@noop {} {\bibfield  {journal} {\bibinfo  {journal}
  {Zeitschrift f{\"u}r Physik A Hadrons and nuclei}\ }\textbf {\bibinfo
  {volume} {209}},\ \bibinfo {pages} {93} (\bibinfo {year} {1968})}\BibitemShut
  {NoStop}%
\bibitem [{\citenamefont {Fefferman}\ and\ \citenamefont
  {Weinstein}(2014)}]{fefferman2014wave}%
  \BibitemOpen
  \bibfield  {author} {\bibinfo {author} {\bibfnamefont {C.~L.}\ \bibnamefont
  {Fefferman}}\ and\ \bibinfo {author} {\bibfnamefont {M.~I.}\ \bibnamefont
  {Weinstein}},\ }\bibfield  {title} {\bibinfo {title} {Wave packets in
  honeycomb structures and two-dimensional dirac equations},\ }\href@noop {}
  {\bibfield  {journal} {\bibinfo  {journal} {Communications in Mathematical
  Physics}\ }\textbf {\bibinfo {volume} {326}},\ \bibinfo {pages} {251}
  (\bibinfo {year} {2014})}\BibitemShut {NoStop}%
\bibitem [{\citenamefont {Ammari}\ \emph {et~al.}(2020)\citenamefont {Ammari},
  \citenamefont {Hiltunen},\ and\ \citenamefont {Yu}}]{ammari2020high}%
  \BibitemOpen
  \bibfield  {author} {\bibinfo {author} {\bibfnamefont {H.}~\bibnamefont
  {Ammari}}, \bibinfo {author} {\bibfnamefont {E.~O.}\ \bibnamefont
  {Hiltunen}},\ and\ \bibinfo {author} {\bibfnamefont {S.}~\bibnamefont {Yu}},\
  }\bibfield  {title} {\bibinfo {title} {A high-frequency homogenization
  approach near the dirac points in bubbly honeycomb crystals},\ }\href@noop {}
  {\bibfield  {journal} {\bibinfo  {journal} {Archive for Rational Mechanics
  and Analysis}\ }\textbf {\bibinfo {volume} {238}},\ \bibinfo {pages} {1559}
  (\bibinfo {year} {2020})}\BibitemShut {NoStop}%
\bibitem [{\citenamefont {Wallace}(1947)}]{wallace1947band}%
  \BibitemOpen
  \bibfield  {author} {\bibinfo {author} {\bibfnamefont {P.~R.}\ \bibnamefont
  {Wallace}},\ }\bibfield  {title} {\bibinfo {title} {The band theory of
  graphite},\ }\href@noop {} {\bibfield  {journal} {\bibinfo  {journal}
  {Physical review}\ }\textbf {\bibinfo {volume} {71}},\ \bibinfo {pages} {622}
  (\bibinfo {year} {1947})}\BibitemShut {NoStop}%
\bibitem [{\citenamefont {Ablowitz}\ \emph {et~al.}(2009)\citenamefont
  {Ablowitz}, \citenamefont {Nixon},\ and\ \citenamefont
  {Zhu}}]{ablowitz2009conical}%
  \BibitemOpen
  \bibfield  {author} {\bibinfo {author} {\bibfnamefont {M.~J.}\ \bibnamefont
  {Ablowitz}}, \bibinfo {author} {\bibfnamefont {S.~D.}\ \bibnamefont
  {Nixon}},\ and\ \bibinfo {author} {\bibfnamefont {Y.}~\bibnamefont {Zhu}},\
  }\bibfield  {title} {\bibinfo {title} {Conical diffraction in honeycomb
  lattices},\ }\href@noop {} {\bibfield  {journal} {\bibinfo  {journal}
  {Physical Review A}\ }\textbf {\bibinfo {volume} {79}},\ \bibinfo {pages}
  {053830} (\bibinfo {year} {2009})}\BibitemShut {NoStop}%
\bibitem [{\citenamefont {Peres}(2009)}]{peres2009scattering}%
  \BibitemOpen
  \bibfield  {author} {\bibinfo {author} {\bibfnamefont {N.}~\bibnamefont
  {Peres}},\ }\bibfield  {title} {\bibinfo {title} {Scattering in
  one-dimensional heterostructures described by the dirac equation},\
  }\href@noop {} {\bibfield  {journal} {\bibinfo  {journal} {Journal of
  Physics: Condensed Matter}\ }\textbf {\bibinfo {volume} {21}},\ \bibinfo
  {pages} {095501} (\bibinfo {year} {2009})}\BibitemShut {NoStop}%
\bibitem [{\citenamefont {Carminati}\ and\ \citenamefont
  {Schotland}(2021)}]{carminati2021principles}%
  \BibitemOpen
  \bibfield  {author} {\bibinfo {author} {\bibfnamefont {R.}~\bibnamefont
  {Carminati}}\ and\ \bibinfo {author} {\bibfnamefont {J.~C.}\ \bibnamefont
  {Schotland}},\ }\href@noop {} {\emph {\bibinfo {title} {Principles of
  Scattering and Transport of Light}}}\ (\bibinfo  {publisher} {Cambridge
  University Press},\ \bibinfo {year} {2021})\BibitemShut {NoStop}%
\end{thebibliography}%

\end{document}